\documentclass[11pt]{article} 
\usepackage{amssymb}
\usepackage{amsmath}
\usepackage{graphicx}
\usepackage{bm}
\newcommand{\mathsym}[1]{{}}

\usepackage{graphicx}
\usepackage{rotating}
\usepackage{color}
\usepackage{cancel}
\usepackage{hyperref}
\usepackage{doi}

\newcommand{\bra}{\begin{array}}
\newcommand{\era}{\end{array}}
\newcommand{\beq}{\begin{equation}}
\newcommand{\eeq}{\end{equation}}
\newcommand{\beqar}{\begin{eqnarray}}
\newcommand{\eeqar}{\end{eqnarray}}

\newcommand{\be}{\begin{equation}}
\newcommand{\ee}{\end{equation}}
\newcommand{\bea}{\begin{eqnarray}}
\newcommand{\eea}{\end{eqnarray}}
\newcommand{\bd}{\begin{displaymath}}
\newcommand{\ed}{\end{displaymath}}

\hypersetup{colorlinks=true,linkcolor=blue,urlcolor=blue,citecolor=blue}
\usepackage{orcidlink}
\usepackage{cite}

\numberwithin{equation}{section}

\begin{document}

\vspace{20pt}

\begin{center}

{\Large \bf Modified Kerr black holes 
surrounded by dark matter spike
\medskip
 }
\vspace{15pt}

S. Capozziello\orcidlink{0000-0003-4886-2024}
\footnote{capozziello@na.infn.it}${}^{,a,b,c} $,  
S. Zare\orcidlink{0000-0003-0748-3386}
\footnote{szare@uva.es}${}^{,d}$, 
L.M. Nieto\orcidlink{0000-0002-2849-2647}
\footnote{luismiguel.nieto.calzada@uva.es, Corresponding author}${}^{,d}$, and 
H. Hassanabadi\orcidlink{0000-0001-7487-6898}
\footnote{hha1349@gmail.com}${}^{,e}$

\vspace{15pt}
{\sl ${}^{a}$
Dipartimento di Fisica ``E. Pancini", Universit\`a degli Studi di Napoli, ``Federico II" \\
	Complesso Universitario Monte S. Angelo,
	Via Cinthia 9 Edificio G, 80126 Napoli, Italy
}

{\sl ${}^{b}$
Istituto Nazionale di Fisica Nucleare (INFN),
	Sezione di Napoli Complesso Universitario Monte S. Angelo,
	Via Cinthia 9 Edificio G, 80126 Napoli, Italy
 }

{\sl ${}^{c}$ Scuola Superiore Meridionale, Largo San Marcellino 10, 80138 Napoli, Italy
 }

{\sl ${}^{d}$ Departamento de F\'{\i}sica Te\'orica, At\'omica y Optica and Laboratory for Disruptive \\ Interdisciplinary Science (LaDIS), Universidad de Valladolid, 47011 Valladolid, Spain
 }

{\sl ${}^{e}$ Department   of   Physics,   University   of   Hradec   Kr\'{a}lov\'{e}, Rokitansk\'{e}ho   62,   500   03   Hradec   Kr\'{a}lov\'{e},   Czech Republic
 }

\vspace{15pt}
\end{center}

\begin{abstract}
We study supermassive black holes (SMBH), surrounded by a dark matter (DM) spike, that can be found at the centers of  Milky Way and $\text{M87}$ galaxies and are accompanied by a specific kind of topological defect.   The investigation is developed within the framework of Bumblebee Gravity with a global monopole (BGGM).
The dark matter spike is described by a power-law density profile.
Our main objective is to assess how the background arising from spontaneous Lorentz symmetry breaking and the presence of a global monopole influence the properties of the Kerr BH within the region affected by the spike.
Using a spherically symmetric static BH with BGGM properties as the seed metric, 
we construct a non-rotating spacetime with a DM spike, resulting in a BGGM-motivated Schwarzschild-like BH by solving the modified Tolman-Oppenheimer-Volkoff equations (TOV).
Next, we extend this approach to the case of a rotating spacetime 
resulting in the BGGM-motivated Kerr-like BH (BGMKLBH). 
This approach allows us to explore the spacetime structure, and the BGMKLBH shadows.
Then, using available observational data for the DM spike density and considering the effects of BGGM on $\text{Sgr A}^{*}$ and $\text{M87}^{*}$ SMBHs, we analyse the shapes of their shadows and put constraints on the BGGM parameter. 
Thus, we infer that the BGMKLBHs could be reliable candidates for the astrophysical BHs.
\end{abstract}

 \vspace{10pt}


\section{Introduction} \label{sec1}
Modified gravity theories, driven by diverse motivations at ultraviolet regime, such as probing fundamental physics in strong gravitational fields (GFs), or, at infrared regimes,  addressing cosmological and astrophysical issues as DM and dark energy, offer a promising avenue for advancing our understanding of gravity. These theories hold the potential to unveil new insights into the deep nature of gravity and the structure of the Universe, making them a vibrant focus point in today  cosmology and astrophysics \cite{CapozzielloRP2011,CaiRPP2016,NojiriPR2017,JusufiEPJC2022,LambiaseJCAP2023,LambiaseApJ2020,CapozzielloJCAP2017,CapozzielloPRD2014} 
In particular, black holes (BHs), the most fascinating objects predicted by general relativity (GR), have garnered significant interest in the realm of astrophysics. Recent advancements, such as the imaging of $\text{M87}^{*}$ \cite{EventHorizonL1,EventHorizonL2,EventHorizonL3,EventHorizonL4,EventHorizonL5,EventHorizonL6} and $\text{Sgr A}^{*}$ \cite{EventHorizonL12,EventHorizonL13,EventHorizonL14,EventHorizonL15,EventHorizonL16,EventHorizonL17} SMBHs by the EHT collaboration, along with the detection of X-rays  \cite{FabianMNRAS1989} and gravitational waves \cite{AbbottPRX2016}, have solidified the belief that, at the centers of galaxies, lie entities governed solely by gravity.
Furthermore, BHs serve as astrophysical laboratories, enabling the exploration of  theories of gravity and cosmology through a multitude of strong-field phenomena, including the  BH shadows in the presence of DM distributions \cite{Bardeen1973}.
Likewise, the Kerr hypothesis proposes that astrophysical BHs possess unique features that can be described by the Kerr metric \cite{KerrPRL1963}. 
This metric is the only asymptotically flat, axially symmetric, and unique stationary vacuum solution of the Einstein equations \cite{AnjumPoDU2023,KumarApJ2020-1}.
Recent EHT images of the SMBHs $\text{M87}^{*}$ and  $\text{Sgr A}^{*}$ probed that the seen shadows are compatible with what would be expected from a Kerr BH in the context of GR.
Images and shadows \cite{Luminet1979,FalckeApJ2000,HiokiPRD2009,FengEPJC2020,Cunha2015PRL,WeiJCAP2013,PerlickPR2022,JohnsonSA2020,SolankiPRD2022,GrenzebachPRD2014,TsupkoPRD2017,VagnozziPRD2019,BambiPRD2019,HouJCAP2018,KonoplyaPLB2019,ShaikhPRD2019,KhodadiJCAP2020,AfrinMNRAS2021,AfrinMNRAS2023,ParbinPoDU2023,UniyalPoDU2023,OvgunJCAP2018,PantigFP2022,LiPRD2020,WangJCAP2023,Addazi1,Addazi2,PsaltisPRL2020,JusufiEPJC2023,JusufiMNRA2021,Theodosopoulos2023} resulting from the gravitational lensing (GL) of light  \cite{VirbhadraPRD2000,VirbhadraPRD2022,BozzaGRG2001,BozzaPRD2002,IslamJCAP2020,KuangPRD2022,HallaPRD2023} provide crucial insights into the GFs surrounding Kerr BHs, helping to reveal their intrinsic characteristics.
The BH event horizon generates an extremely intense GF, which influences the surrounding spacetime geometry. This, in turn, can lead to the development of unstable circular photon orbits, referred to as unstable light rings (or a photon sphere in the case of spherically symmetric, static BHs).
These phenomena result in the significant bending of photons, leading to pronounced GL effects of a remarkable scale.
For photons on such unstable orbits, even a small perturbation can send them off to a distant observer or absorb them by the BH. 
Therefore, it appears that the unstable light rings and the BH event horizon will give rise to a distinctive shadow-like image of photons from surrounding light sources or radiation from an accretion flow surrounding the BH -- a darker area set against a brighter background.
As the shadow silhouette relates to the apparent shape of the unstable photon orbits, as perceived by a distant observer, it is governed only by the spacetime metric, unlike the intensity map of an image, which is dependent on the specifics of the photon emission mechanisms. 
Thus, strong lensing images and shadows provide us with a unique opportunity to evaluate both GR and alternative gravity theories \cite{PsaltisPRL2020}. Hence, it is key to continue analytical efforts to determine the shapes of shadows cast by BHs and BH mimickers in various gravity and astrophysical theories. Shadow images can reveal details regarding a variety of astrophysical issues, such as matter accretion around BHs and the distribution of DM in galaxies' centers \cite{KonoplyaPLB2019,HouJCAP2018,PantigFP2022,JusufiEPJC2023,JusufiMNRA2021,XuJCAP2018-2,JusufiPRD2019,JusufiEPJC2020354,PantigJCAP2022,CapozzielloJCAP2023,NampalliwarAJ2021}. 
	
Black holes in our Universe could potentially experience influences from astronomical surroundings, including the presence of DM in their vicinity \cite{XuJCAP2021,DaghighApJ2022,XuJCAP2018,XuPRD2020,PantigEPJC2022,GondoloPRL1999,SadeghianPRD2013,GorchteinPRD2010,LacroixPRD2015,LacroixPRD2017,NishikawaPRD2019,KavanaghPRD2020}.
In addition to investigating BH shadows within the framework of modified gravity, it is noteworthy that examining these shadows in the presence of DM and dark energy holds particular significance. This is due to the overwhelming dominance of DM (constituting $27\%$ of the Universe) and dark energy (making up $68\%$), while baryonic matter's contribution is relatively minor (comprising only $5\%$ of the total mass-energy of the Universe) according to the Standard Model (SM) of Cosmology.
The cosmic microwave background radiation, baryon acoustic oscillations, spiral galaxy rotation curves, and mass-luminosity ratios of elliptical galaxies provide compelling evidence for the existence of a surrounding DM halo that extends into the intergalactic medium \cite{BullockARAA2017,PlanckAA2015,GuoMNRAS2016}. 
This halo's DM density profile could be crucial in establishing the real geometry of spacetime around the galactic center \cite{XuJCAP2021,NampalliwarAJ2021,CapozzielloJCAP2023,DaghighApJ2022}. This DM distribution around $\text{M87}^{*}$ and  $\text{Sgr A}^{*}$ SMBHs, in particular, is highly relevant to verify and further constrain, the predictions of GR and any alterations beyond GR. Additionally, it will aid in identifying the DM candidates.  It is worth wondering how the DM around the BHs affects the spacetime associated with these BHs. 
Based on the adiabatic approximation, several models for the spacetime metric around a static and spherically symmetric BH with a DM halo have been presented \cite{BoshkayevMNRAS2020,BoshkayevApJ2022,BoshkayevMNRAS2021,BoshkayevMNRAS2019},  namely the Navarro-Frenk-White (NFW) profile \cite{NavarroAJ1996}. 
	
Some prior results \cite{GondoloPRL1999,NavarroAJ1996,QuinlanAJ1995,UllioPRD2001} show that the presence of a central BH causes an accumulation of DM particles in its strong gravitational potential, generating a spike distribution towards the BH horizon. The BH gravitational field causes the DM density to substantially increase by many orders of magnitude. In turn, the intensity of gamma-ray radiation close to the BH will significantly rise if DM particles are capable of annihilating into gamma-ray radiation. Hence, this provides an opportunity to find the DM annihilation signal \cite{XuJCAP2021}.
	
\begin{figure}[htb]
\centering 
\includegraphics[width=.3\textwidth]{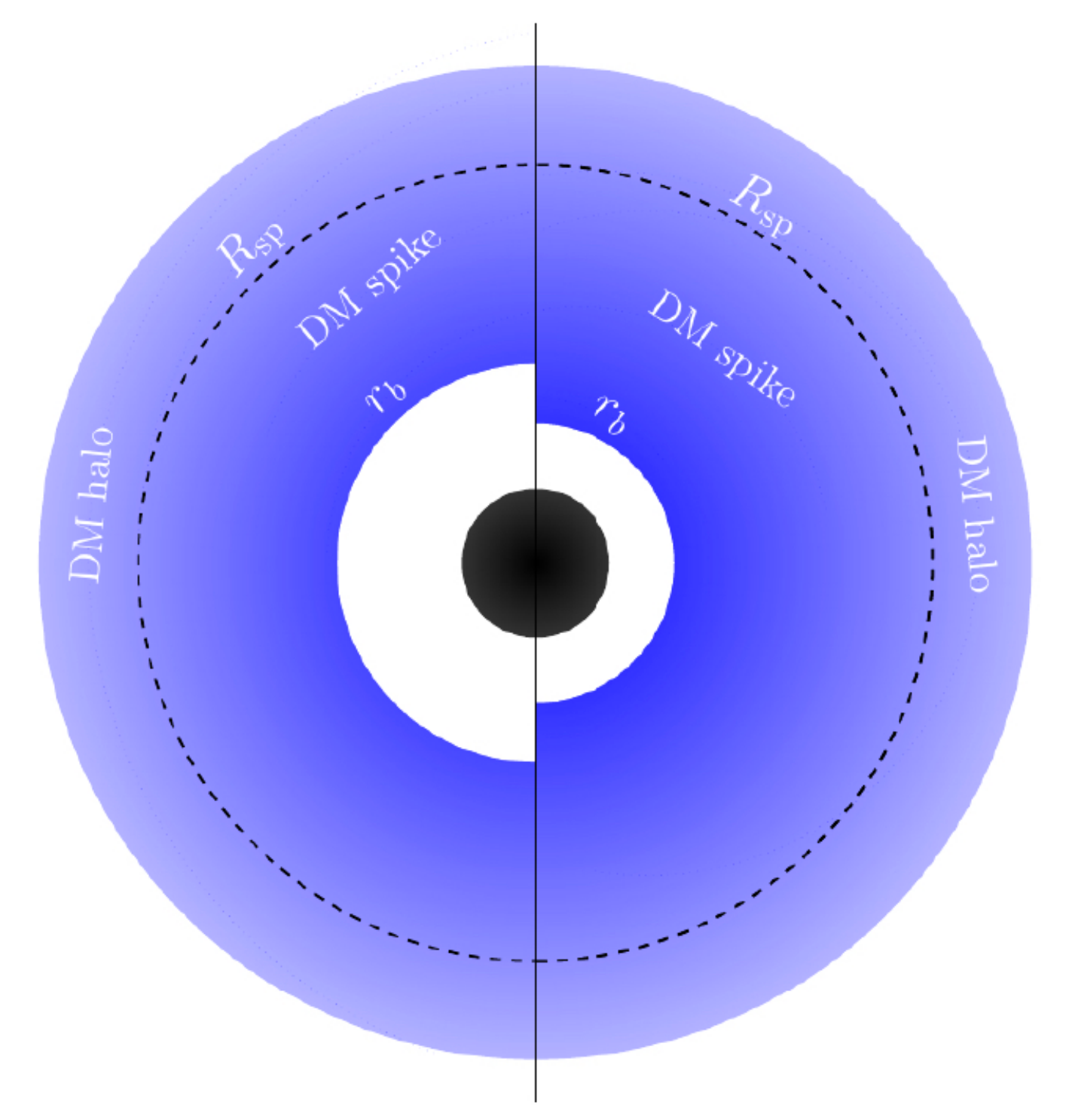}
\caption{
Schematic plot of the galactic central region with an SMBH and a DM spike distribution in the region $r\in[r_{\text{b}}, R_{\text{sp}}]$, where the inner and outer edges of the spike region are given by $r_{\text{b}}$ and $R_{\text{sp}}$. 
Comparison of the density distribution of DM around Schwarzschild BHs, showing a modified model by Sadeghian-Ferrer-Will (SFW) (right half of the panel) incorporating general relativistic corrections, which predicts the DM density to begin at $r = 2R_{s}$, in contrast to the Gondolo and Silk (GS) consideration (left half of the panel), where the density was expected to start at $r = 4R_{s}$.
}\label{fig:Sec21} 
\end{figure}

A BH has a profound effect on the density distribution of DM \cite{ZhaoPRD2023}. An early groundbreaking paper \cite{GondoloPRL1999} determined the distribution of cold DM in the vicinity of galactic centers using a Newtonian approach. 
The BH accretion results in the formation of a density cusp, known as DM spike, characterized by a density profile $\rho\sim r^{-\gamma_{\text{sp}}}$, where $2.25\leq\gamma_{sp}\leq2.5$. For spherically symmetric BHs, the density reaches its peak at approximately $r \sim 4R_{s}$, with $R_{\text{s}}$ representing the Schwarzschild radius. Below this point, that is, $r = 4R_{s}$, there is a rapid decline in DM density, as particles either annihilate or fall into the BH.
When considering relativistic modifications \cite{SadeghianPRD2013}, the density profile exhibits similar traits, albeit with a change in the cutoff radius, now occurring at $r = 2R_{s}$ (see Figure~\ref{fig:Sec21}).

At the heart of the Milky Way and $\text{M87}$ galaxies, we plan to study SMBHs, which are surrounded by a DM spike accompanied by a unique kind of topological defect.  
Our investigation is rooted in Bumblebee Gravity (BG) coupled with a global monopole (GM), with a particular focus on analyzing the trace of the model parameters (the BGGM parameters $\ell$ and $\tilde{\mu}$). 
	
Spontaneous symmetry breaking, a fundamental concept in particle physics, can manifest in two distinct forms: as an internal symmetry or as a symmetry linked to spacetime transformations. The spontaneous breaking of internal symmetries gives rise to the formation of global topological defects \cite{GulluAP2022}.
One type of stable topological defect is a monopole \cite{BarriolaPRL1989,VilenkinNPB1989}. 
The origin of inflation may come from monopoles that are created when gauge-symmetry breaks during early Universe phase transitions. However, at phase transitions in the Universe, global monopoles arise from a global symmetry breaking of global $O(3)$ symmetry into $U(1)$ \cite{GulluAP2022,BarriolaPRL1989,VilenkinNPB1989}.
The BG model is an extension of the standard framework of GR that allows for the spontaneous breaking of Lorentz symmetry (LS) via a non-zero vacuum expectation value (VEV) of the bumblebee vector field, denoted as $B_{\mu}$, achieved through an appropriate potential. This model serves as a prominent example of a theory showcasing Lorentz violation, originating from a single vector $B_{\mu}$ acquiring a non-zero VEV. It stands out as one of the simplest field theories manifesting spontaneous Lorentz and diffeomorphism violations \cite{Kostelecky,BluhmPRD2005,KosteleckPotting,KosteleckyTasson,CasanaPRD2018,GogoiJCAP2022,Zhang2023,GulluAP2022,GomesAP2020,KhodadiEPJC2023,Khodadiarxiv2022,JhaEPJC2021,JhaJCAP2021}. 
In this case, LS breaking emerges due to the presence of a potential with a functional form that has a minimum, leading to the violation of $U(1)$ symmetry.  
The concept of the bumblebee formalism draws inspiration from string theory, where it is postulated that tensor fields can acquire VEV and thus lead to the spontaneous breaking of LS \cite{KosteleckyTasson}. Recent advancements in this domain include the derivation of the exact solution for the Schwarzschild Bumblebee BH \cite{CasanaPRD2018,GulluAP2022,GogoiJCAP2022}.

In this work, we evaluate the effects of such an extended gravity model background resulting from spontaneous LS breaking and a GM (aptly considered as the BGGM model) on horizons, static limit surfaces (SLS)s, ergoregions, and shadow silhouettes of the Kerr-like BHs in the spike-affected region. The progress in testing the BGGM model through observations is greatly impeded by the lack of both rotating and non-rotating BH models immersed in a DM spike and subject to BGGM effects. 
We begin our inquiry to tackle this problem by taking as the seed metric a spherically symmetric static BH with BGGM characteristics. From there, we construct the Schwarzschild-like non-rotating spacetime with a DM spike, which we introduce as BGGM-motivated Schwarzschild-like BH (BGMSLBH) spacetimes.
To construct the BGMSLBH spacetimes, we start with the power-law density profile as originally proposed by GS. Then, we solve the modified TOV equation, approximating the integral in the leading order for the spike density. Our approach involves the critical condition of matching the inner BH spacetime with the outer region, specifically employing the condition denoted as ${\rm{f}}(r_{\text{b}}) = e^{2\chi(r_{\text{b}})} = 1-\frac{2M_{\text{BH}}}{r_{\text{b}}}$, in line with the methodology presented by Nampalliwar et al. Ref. \cite{NampalliwarAJ2021}. This process yields the corresponding metric components, where $\text{f}(r)\neq\text{g}(r)$.
We then use the modified Newman-Janis (NJ) algorithm to extend this approach to the case of a rotating spacetime, yielding BGMKLBH
\footnote{We refer to Kerr BHs immersed in DM spikes and subjected to BGGM effects as 'rotating BGMSLBH' or simply 'BGMKLBH' for the sake of clarity and conciseness in this study.} 
spacetimes. Next, we investigate the BGMKLBH's horizons, SLSs, ergoregions, and shadow silhouettes.
	
Besides, we intend to determine whether the EHT findings for $\text{M87}^{*}$ and  $\text{Sgr A}^{*}$ can shed light on the BGGM model in the DM spike-affected region and constrain the BGGM parameter. 
The EHT has provided observational results regarding the mass and distance of both $\text{M87}^{*}$ and  $\text{Sgr A}^{*}$ while setting constraints on their shadow observables. 
By modeling BGMKLBHs as $\text{M87}^{*}$ and  $\text{Sgr A}^{*}$, we aim to assess their potential as candidates for SMBHs and to establish astrophysical bounds on the BGGM parameter through direct analysis of BH shadows. 
Additionally, we seek to ascertain if the BGMKLBH can offer robust constraints on the BGGM parameter for the $\text{M}87^{*}$ and $\text{Sgr A}^{*}$ SMBHs.
	
The paper is organized as follows: In Section 2, we begin by taking a seed static, spherically symmetric BGGM BH metric. We then introduce the DM spike profile and proceed to calculate the spacetime metric for the corresponding Schwarzschild-like BH, which are surrounded by a DM distribution in the spike-affected region.
In Section 3, we calculate the normalization parameter, $\rho_{0}$, and related parameters for the DM spike profiles at the centers of both  Milky Way and  $\text{M87}$ galaxies.
In Sections 4 and 5, we derive the BGMKLBHs using the modified NJ method and we investigate how the BGGM model affects this deformed Kerr-like BH horizons, SLSs, ergoregions and shadows in the spike region.
In Section 6, we constrain the BGGM parameter using EHT shadow observations of $\text{M87}^{*}$ and $\text{Sgr A}^{*}$ at inclination angles of $17^{\circ}$ and $46^{\circ}$.
In Section 7, we briefly present our results and draw our conclusions.
Throughout the paper, we adopt  natural units, in which $G$, $c$, and $\hbar$ are all equal to $1$.

\section{The Schwarzschild Bumblebee BH with a global monopole in DM spike
	\label{sec2}}
Inspired by the investigation of the central BH's characteristics within a realistic framework surrounded by a DM distribution and significantly influenced by the breaking of LS in the presence of a GM, we will construct the new spacetime metric around a static and spherically symmetric BGGM-motivated Schwarzschild BH immersed in a DM spike. To achieve this, we begin with a static, spherically symmetric BGGM BH seed metric given by \cite{CasanaPRD2018,GogoiJCAP2022,GulluAP2022,GomesAP2020}
\begin{equation}\label{Bumbelbeemetric1}
	\begin{split}
		ds^2= -\left(w-\frac{2M_{\text{BH}}}{r}\right) dt^{2}+q\left(w-\frac{2M_{\text{BH}}}{r}\right)^{-1}dr^{2}+r^{2}d\Omega^{2}.
	\end{split}
\end{equation}
Here, the constants $w$ and $q$ are defined as $w=1-\tilde{\mu}$ and $q = 1 + \ell$,
where the LS breaking parameter, denoted by $\ell$, takes values in the range  $(0,1)$ and the GM term is defined as $\tilde{\mu} = -\eta^{2}$, where $\eta$ is a constant term corresponding to the GM charge, as well as the line element of unit two-sphere is given by $d\Omega^{2} = d\theta^{2}+\sin^{2}\theta d\varphi^{2}$.
We note that metric \eqref{Bumbelbeemetric1} yields the standard spherically symmetric solution with LS breaking when $\eta$ equals zero. Moreover, for $\eta$ equals zero and $\ell$ approaching zero,  the standard Schwarzschild metric is recovered.
Furthermore, in the presence of BG,  which is responsible for the effects of LS breaking, and the GM, singularities exist at $r = 2M_{\text{BH}}/w$ and $r = 0$.
In this scenario, the event horizon of the BH is located at $g_{tt}(r_{0}) = 0$, resulting in $r_{0} = 2M_{\text{BH}}/w$. It is seen that this value is independent of $\ell$ and only relies on $\tilde{\mu}$.
Now, let us proceed with the calculation of the Kretschmann scalar associated with the BGGM metric \eqref{Bumbelbeemetric1} to analyze the nature of the singularities. The Kretschmann scalar is given by:
\begin{equation}\label{KretschmannBBHGM}
		K_{\text{Kretschmann}}=\mathcal{R}_{\alpha\beta\mu\nu}\mathcal{R}^{\alpha\beta\mu\nu} =\frac{48M_{\text{BH}}^{2}}{q^{2}r^{6}}+\frac{4\left(q-w\right)}{q^{2}r^{5}}\left(\left(q-w\right)r+4M_{\text{BH}}\right).
\end{equation}
In the limit of  $\ell$ and $\tilde{\mu}$ approaching to zero, this Kretschmann scalar \eqref{KretschmannBBHGM}, reduces to $48M_{\text{BH}}^{2}/r^{6}$, corresponding to the Kretschmann scalar of the standard Schwarzschild BH. 
As observed in Eq. \eqref{KretschmannBBHGM}, the Kretschmann scalar at $r = 0$ exhibits a divergence, indicating the presence of a physical singularity. However, for $r_{0} = 2M_{\text{BH}}/w$, the Kretschmann scalar is finite, specifically $w^{4}(q^{2}+2w^{2})/4M_{\text{BH}}^{4}q^{2}$. This implies that the singularity at the event horizon can be eliminated through a coordinate transformation.

\subsection{Dark matter modeling in general relativity}
\begin{figure}[htb]
	\centering 
	\includegraphics[width=.5\textwidth]{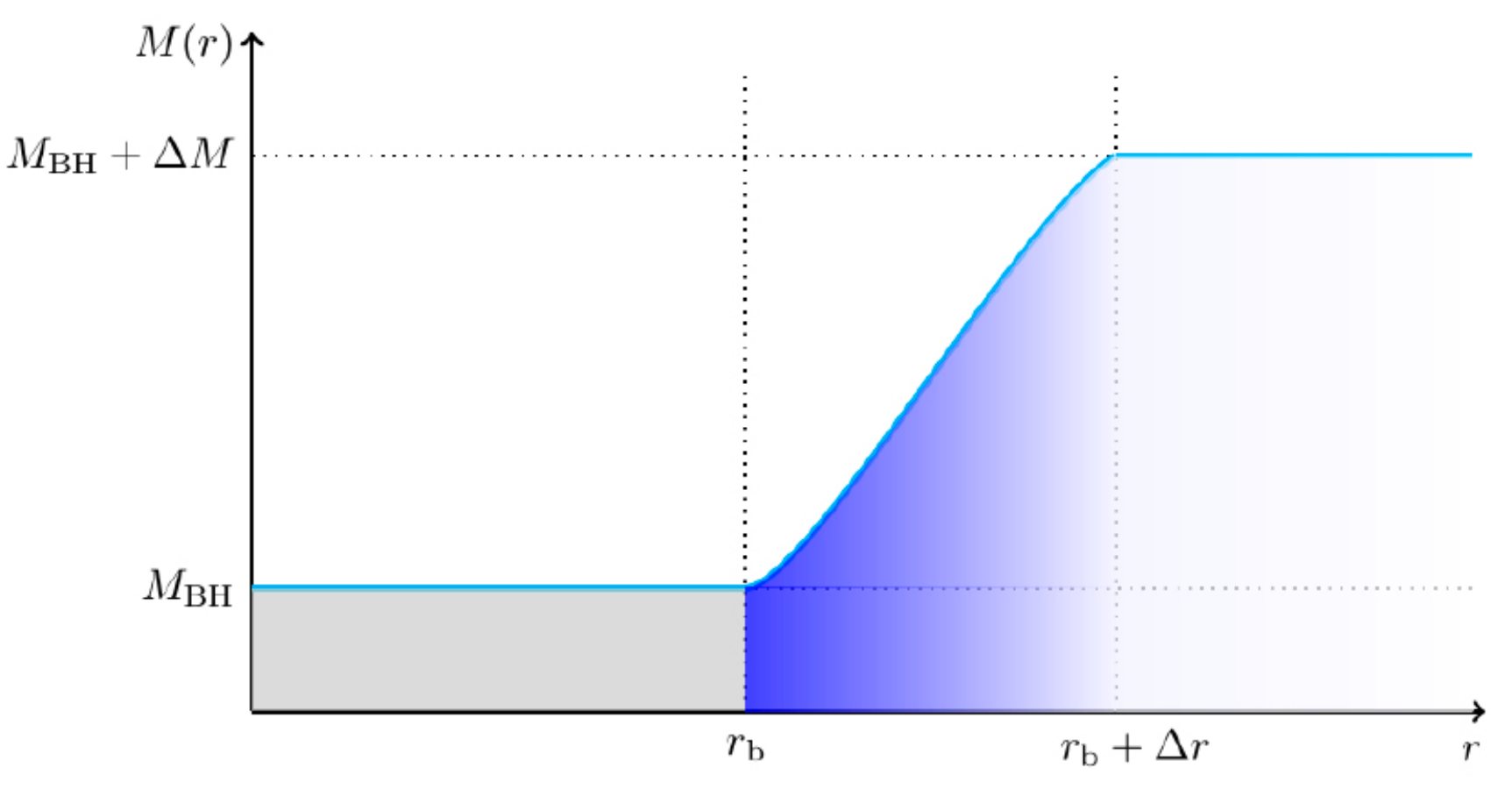}
	\caption{\label{fig:MassBHDM} 
		Schematic plot of the galactic central region with an SMBH and a DM distribution in the region $r\in[r_{\rm{b}}, r_{\rm{b}}+\Delta r]$, where the inner and outer edges of the DM region are given by $r=r_{\rm{b}}\geq2M_{\rm{BH}}$ and $r = r_{\rm{b}}+\Delta r$. The event horizon is identified at $r_{+} = 2M_{\rm{BH}}$.
		The shaded region illustrates a steep DM density profile around a BH, resulting from the BH's gravitational pull. This occurs as DM is adiabatically drawn inward, causing a significant increase in density near the BH.
	}
\end{figure}

As reported in  \cite{KonoplyaPLB2019}, some approaches exist for modeling supermassive BHs at the centers of galaxies, based on current cosmological observations \cite{BoshkayevMNRAS2020,BoshkayevApJ2022,BoshkayevMNRAS2021,BoshkayevMNRAS2019}. Here, we adopt a more agnostic approach, using the fact that DM has mass, which can be treated as an additional effective mass in the collective mass function  $m(r)$, embedded within the most general 4D spherically symmetric static metric:
\begin{equation}\label{GeneralMetric}
	ds^{2}=-f(r)dt^{2}+\frac{1}{f(r)}dr^{2}+r^{2}(d\theta^{2}+\sin^{2}\theta d\varphi^{2}), 
\end{equation}
with $f(r) = 1-2m(r)/r$, where the collective mass function  $m(r)$ is defined as
\begin{equation}\label{CollMassFunc}
	m(r)=
	\begin{cases}
		M_{\rm{BH}}, & \text{for \quad} r \leq r_{\rm{b}};  \\
		M_{\rm{BH}}+ W(r)\Delta M, & \text{for \quad} r_{\rm{b}}  \leq r \leq r_{\rm{b}}+\Delta r;\\
		M_{\rm{BH}}+ \Delta M, & \text{for \quad} r > r_{\rm{b}}+\Delta r,
	\end{cases}
\end{equation}
in which $W(r) = (3-2(r-r_{\rm{b}})/\Delta r)(r-r_{\rm{b}})^{2}/\Delta r^{2}$ is defined as a radial function to ensure the continuity of the mass function and its first derivative with respect to $r$ (refer to Figure~\ref{fig:MassBHDM}).
In this framework for modeling DM within GR, it is anticipated that the DM density begins at $r_{\rm{b}} > r_{+} = 2M_{\rm{BH}}$ and extends to $r_{\rm{b}} + \Delta r$. Here, $\Delta M$ represents the mass of the DM distribution, with $\Delta M > 0$ indicating positive DM mass-energy density and $\Delta M < 0$ signifying a negative density. In this study, we focus exclusively on the positive case, with $\Delta r$ representing the thickness of the DM distribution.

\begin{figure}[htb]
	\centering 
	\includegraphics[width=.45\textwidth]{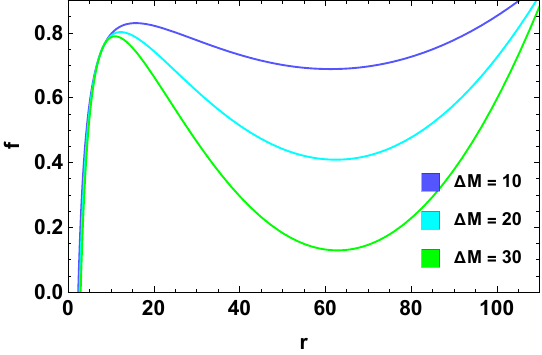}
	\hfill
	\includegraphics[width=.45\textwidth]{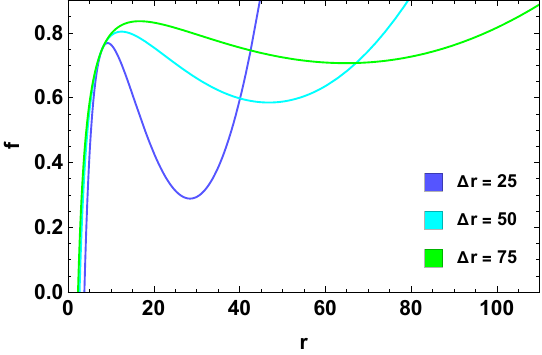}
	\caption{\label{fig:MassBHDM2} 
		Plot of the metric function versus radial coordinate for $\Delta r= 70$ showing variations in  $\Delta M $ (first panel) and for $\Delta M = 10M_{\rm{BH}}$ with varying $\Delta r $ (second panel), with parameters set to $M_{\rm{BH}} = 1$ and $r_{\rm{b}} = 8$.
	}
\end{figure}

Figure~\ref{fig:MassBHDM2} illustrates the behavior of the metric function $f(r)$ in the DM model for various combinations of $\Delta M$ and $\Delta r$, while holding certain values of $r_{\rm{b}}$ and $\Delta r$ constant. 
The admissible values of $\Delta M$ and $\Delta r$ are subject to constraints. If $\Delta M$ is excessively large, it effectively raises the total mass of the BH, consequently expanding the radius of the event horizon. Conversely, reducing the thickness of the DM distribution also leads to an increase in the event horizon radius.

\subsection{The modified TOV equations with DM spike}
We proceed by considering a BH positioned at the center of a DM halo. The BH possesses a mass denoted as $M_{\text{BH}}$. Initially, the DM halo exhibits a power-law density profile in close proximity to the galactic center, expressed as $\rho_{\text{DM}}(r) \simeq \rho_{0}(r_{0}/r)^{\gamma}$. Here, $\gamma$ represents the power-law index, while $\rho_{0}$ and $r_{0}$ serve as the parameters characterizing the halo. In a study by Gondolo and Silk \cite{GondoloPRL1999}, 
it was demonstrated that the formation of a DM spike follows an adiabatic process, resulting in a density profile $\rho^{\text{sp}}_{\text{DM}}$ equal to
\begin{equation}\label{densityDMS1}
	\rho^{\text{sp}}_{\text{DM}}(r) = \rho_{\text{sp}}\left(\frac{R_{\text{sp}}}{r}\right)^{\gamma_{\text{sp}}},
\end{equation}
where $\rho_{\text{sp}} = \rho_{0}\left(R_{\text{sp}}/r_{0}\right)^{-\gamma}$ refers to the DM density, while 
$R_{\text{sp}} = \mathcal{N}_{\gamma} r_{0}(M_{\text{BH}}/\rho_{0} r_{0}^{3})^{1/(3-\gamma)}$ corresponds to the spike radius (see Figure~\ref{fig:Sec21}). Both $\rho_{\text{sp}}$ and $R_{\text{sp}}$ are associated with the outer edge of the spike region \cite{SadeghianPRD2013,GorchteinPRD2010,LacroixPRD2015,LacroixPRD2017,NishikawaPRD2019, KavanaghPRD2020,NampalliwarAJ2021,CapozzielloJCAP2023,DaghighApJ2022,XuJCAP2021}. 
Here, the quantity $\gamma_{\text{sp}}$ is defined as $(9-2\gamma)/(4-\gamma)$ with $\gamma \in [0,2]$, and $R_{\text{s}}$ represents the Schwarzschild radius of the BH and it  is approximately equal to $2M_{\text{BH}} \simeq 2.95 (M_{\text{BH}}/M_{\odot})$ km. 
By employing this approach, it is possible to demonstrate that the normalization constant $\mathcal{N}_{\gamma}$ is connected to the spike parameters through the relation \cite{DaghighApJ2022}
\begin{equation}
	\mathcal{N}_{\gamma} \simeq \left(\frac{M_{\text{BH}}}{\rho_{\text{sp}} R_{\text{sp}}^{3}}\right)^{3-\gamma}.
\end{equation}
It is worth noting that the density profile of this DM distribution differs from the NFW  density profile, which is based on numerical simulations of collisionless DM particles in galactic halos, with values of $\gamma$ equal to $1$ and $0$, along with BH masses of $M_{\text{BH}}=10^{5} M_{\odot}$ or $M_{\text{BH}}=10^{6} M_{\odot}$ \cite{NishikawaPRD2019}. An intriguing aspect is the significant enhancement of DM density by several orders of magnitude within the spike region. Therefore, it is of great interest to explore the potential implications of this phenomenon on observable signatures originating from central SMBHs, such as $\text{Sgr A}^{*}$ and $\text{M87}^{*}$, situated in spacetime with topological defects, that is, with GMs, within the framework of BG.
The next step is to construct a BGMSLBH metric background that incorporates both the effects of the GM and spontaneous breaking of LS, and governs the trajectories of particles, both massive and massless. 
To do so, we proceed to solve the modified TOV equations \cite{Tolman1939,Oppenheimer1939} arising from the corresponding background, in the DM distribution surrounding the BH. 
Within this framework, to analyse the gravitational signatures of the interacting system between the DM in the spike region and the BH in the BGGM background, we begin with a spherically symmetric static metric that resembles the metric given in Eq. \eqref{Bumbelbeemetric1}, called the BGGM BH metric. The metric can be expressed as 
\begin{equation}\label{Bumbelbeemetric2}
	ds^2 = -e^{2\chi (r)}dt^{2}+e^{2\zeta(r)}dr^{2}+r^{2}d\Omega^{2},
\end{equation}
where $\chi(r)$ and $\zeta(r)$, valid in the region $r_{\text{b}}\leq r \leq R_{\text{sp}}$, represent the sought metric functions. It is noteworthy that choosing $e^{-2\zeta(r)} = \text{g}(r)$ is always possible, resulting in:
\begin{equation}\label{MetCoeffFunc-g1}
	\begin{split}
		\text{g} (r) =q^{-1}\left(w-\frac{2 \,{\rm M}(r)}{r}\right).
	\end{split} 
\end{equation}
Here, we adapt the mass function $m(r)$ given in Eq. \eqref{CollMassFunc} to define $\mathrm{M}(r)$, specifically tailored for the DM spike. Hence, the collective mass function $\mathrm{M}(r)$ can be taken as the combination of the BH mass, and a mass function associated with the presence of DM distribution within the spike region (see Figures~\ref{fig:Sec21} and \ref{fig:MassBHDM}). Thus, the resulting collective mass function can be denoted as \cite{KonoplyaPLB2019}
\begin{equation}\label{totalMass1}
	\mathrm{M}(r)= M_{\text{BH}}+M^{\text{sp}}_{\text{DM}}(r).
\end{equation}
By employing the density profile described in Eq. \eqref{densityDMS1},  we derive the mass function corresponding to the distribution of DM within the spike region, restricted to the range $r_{\text{b}} \leq r \leq R_{\text{sp}}$, which can be expressed as 
\begin{equation}\label{DMSpikeMass1}
		M^{\text{sp}}_{\text{DM}}(r) = 4\pi \int_{r_{\text{b}}}^{r} \rho^{\text{sp}}_{\text{DM}}(\bar{r})\bar{r}^{2}\, d\bar{r}
		= \frac{4\pi  \rho_{\text{sp}}}{3-\gamma_{\text{sp}}}R_{\text{sp}}^{\gamma_{\text{sp}}} \left(r^{3-\gamma_{\text{sp}}}-r^{3-\gamma_{\text{sp}}}_{\text{b}}\right),
\end{equation}			
where $r_{\text{b}}$ represents the inner edge of the spike region. 
Speaking of which, for this scenario, the collective mass function across various regions can be rewritten using Eqs. \eqref{CollMassFunc}, \eqref{totalMass1} and \eqref{DMSpikeMass1} as follows
\begin{equation}\label{totalMass2}
	\mathrm{M}(r)=
	\begin{cases}
		M_{\text{BH}}, & \text{for \quad} r \leq r_{\text{b}};  \\
		M_{\text{BH}}+M^{\text{sp}}_{\text{DM}}(r), & \text{for \quad} r_{\text{b}}  \leq r \leq R_{\text{sp}} ;\\
		M_{\text{BH}}+M_{\text{DM}}, & \text{for \quad} r > R_{\text{sp}}.
	\end{cases}
\end{equation}
The DM mass shell, $M_{\text{DM}}$, is fundamentally a constant mass that relies on its density and position. Specifically, for large scales where $r \geq R_{\text{sp}}$, the spacetime can be seamlessly connected with the density profiles of DM halos that extend beyond the spike region \cite{NampalliwarAJ2021}. Meanwhile, the influence of the DM distribution beyond the spike region on the gravitational signatures originating from the BH at the galactic center is considered negligible.

In this way, the energy-momentum tensors associated with the spacetime metric for such a  interacting system can be expressed as $T^{\mu}_{\,\,\nu} = \text{diag}\left[-\rho(r),P_{r}(r),P_{\theta}(r),P_{\varphi}(r)\right]$.
Thus, these considerations lead to the derivation of the Einstein field equations $G_{\mu\nu} = 8\pi T_{\mu\nu}$ as
\begin{subequations}\label{EinsteinFieldEqs1}
	\begin{align}
		&	8\pi \rho(r) = \frac{q-w+2\text{M}'(r)}{qr^{2}},\label{EinsteinFieldEqs1-1}\\
		&	8\pi P_{r}(r) = -\frac{1}{r^{2}}+\frac{w-\frac{2 {\rm M}(r)}{r}}{q r^{2}}+\frac{2\left(w-\frac{2 {\rm M}(r)}{r}\right)\chi'(r)}{q r},\label{EinsteinFieldEqs1-2}\\
		\nonumber 
		&8\pi P_{\theta}(r)= \frac{{\rm M}(r)-r{\rm M'}(r)}{q\,r}-\frac{1}{q}\left({\rm M}(r)+r\left(-w+{\rm M'}(r)\right) \right)\chi'(r)\\ 		
		&\qquad\qquad \   +\frac{r}{q}\left(r w-2{\rm M}(r)\right)\chi'(r)^{2}
		 +\frac{r}{q}\left(r w-2{\rm M}(r)\right)\chi''(r), \label{EinsteinFieldEqs1-3}\\
		&8\pi P_{\varphi}(r)=8\pi \sin^{2}\theta P_{\theta}(r). \label{EinsteinFieldEqs1-4}
	\end{align}
\end{subequations}
Here, the prime  indicates the derivative of the functions with respect to $r$. When the limits $w\rightarrow 1$ and $q\rightarrow 1$ are applied to the Einstein field equations in Eq. \eqref{EinsteinFieldEqs1}, they reduce to their standard form, which corresponds to the generic ansatz in the standard static and spherically symmetric form $g_{\mu\nu} = \text{diag}(-e^{2\chi(r)},e^{2\zeta(r)},r^{2}, r^{2}\text{sin}^{2}\theta)$.
Then, by combining the equation of state $P_{r}(r) = \omega \rho(r)$ and the GF equation, given in \eqref{EinsteinFieldEqs1-1} and \eqref{EinsteinFieldEqs1-2}, as well as considering the conservation law of the energy-momentum tensor, $T_{;\nu}^{\mu\nu} = 0$, the modified TOV equations can be written as
\begin{subequations}
	\begin{align}
		&\frac{d\chi(r)}{dr} = \,\frac{\text{M}(r)+\left(\frac{q}{2}-\frac{w}{q}\right)r+4\pi r^{3} P_{r}(r)}{r(w r-2\text{M}(r))},\label{TOV2}\\
		&\frac{dP_{r}(r)}{dr} = -\left(\rho(r)+P_{r}(r)\right)\frac{d\chi(r)}{dr},
	\end{align}
\end{subequations}
where $P_{r}(r)$ represents the pseudo-pressure of the DM, which can be defined even for collisionless particles \cite{Binney2008,XuPRD2020}. We adopt the equation of state for the DM spike as $P_{r}(r) = \omega 	\rho^{\text{sp}}_{\text{DM}}(r)$, where $\omega$ can be  a constant.
In order to derive an analytical expression for $\chi(r)$, we are confronted with a complex integral. In this particular case, we employ an asymptotic expansion technique to attain the desired analytical form for $\chi(r)$. 
Here, to simplify the analysis, we consider the case where the DM distribution possesses an equation of state with $\omega = 0$ \cite{JusufiMNRA2021,NampalliwarAJ2021}. Consequently, in Eq. \eqref{TOV2}, the contribution of the term $4\pi r^3 P_{r}(r)$ can be disregarded when compared to $\text{M}(r)$ \cite{DaghighApJ2022}. Thus, Eq. \eqref{TOV2} can be rewritten as
\begin{equation}\label{TOV2-2}
	\frac{d\chi(r)}{dr} = -\frac{1}{2r}+\frac{q}{2}\left(\frac{1}{w r-2\text{M}(r)}\right).
\end{equation}
At this stage, in order to integrate Eq. \eqref{TOV2-2} and determine the metric function $\text{f}(r) = e^{2\chi(r)}$, we have the option to approximate the integral in the leading order with respect to $\rho_{\text{sp}}$. Additionally, using the condition ${\rm {f}}(r_{\text{b}}) = e^{2\chi(r_{\text{b}})} = w-\frac{2{\rm M}_{\text{BH}}}{r_{\text{b}}}$ \footnote{To find further details about the condition, please refer to Ref. \cite{NampalliwarAJ2021}.}, the metric function $\text{f}(r)$ can be expressed as:
{\small
	\begin{equation}\label{MetCoeffFunc-f11}
		\begin{split}
			\text{f}(r)&\simeq r^{-1+\frac{q}{w}}+w-1-\frac{2{\rm M}_{\rm{BH}}}{r}+\frac{8\pi q \rho_{\text{sp}}\, r^{1+\frac{q}{w}}}{w^{2}\left(\gamma_{\text{sp}}-3\right)\left(\gamma_{\text{sp}}-2\right)}\left(\frac{R_{\text{sp}}}{r}\right)^{\gamma_{\text{sp}}}  \\
			&\quad +8\pi\rho_{\text{sp}}\, r^{2}_{\text{b}}\left(\frac{w^{2}r\left(\gamma_{\text{sp}}-3\right) -q\left(\gamma_{\text{sp}}-2\right)r^{-1+\frac{q}{w}}r_{\text{b}}}{w^{2}r\left(\gamma_{\text{sp}}-3\right)\left(\gamma_{\text{sp}}-2\right)}\right)\left(\frac{R_{\text{sp}}}{r_{\text{b}}}\right)^{\gamma_{\text{sp}}}.
		\end{split}
	\end{equation}
	It is observed that this metric is valid in the range $r \in [r_{\rm b}, R_{\rm sp}]$.
}
In the BGMSLBH spacetime described by metric \eqref{Bumbelbeemetric2},  
in the limit of $w\rightarrow 1$ and $q\rightarrow 1$, where the BGGM is absent, metric function \eqref{MetCoeffFunc-f11} simplifies to metric function (18) of Ref. \cite{NampalliwarAJ2021}, as well as when the DM spike disappears  (i.e., $\rho_{\text{sp}} \rightarrow 0$ or $r \rightarrow r_{\text{b}}$), the sought metric functions $\text{f}(r)$ and $\text{g}(r)$ reduces to
\begin{equation}
	\lim \limits_{\substack{%
			\\
			r\to r_{\text{b}}\\
			w \to 1\\
			q \to 1}}   \text{f}(r) = \lim \limits_{\substack{%
			\\
			r\to r_{\text{b}}\\
			w \to 1\\
			q \to 1}}   \text{g}(r) = 1-\frac{2M_{\text{BH}}}{r_{\text{b}}}.
\end{equation}
It is worth noting that the matching condition $\text{f}(r) = \text{g}(r)$ is achieved at the horizon of the BH in the absence of the BGGM. Furthermore, Eq. \eqref{MetCoeffFunc-f11} reveals that the event horizon of the BGMSLBH is influenced by the BGGM effects. Consequently, changes in the variables such as $w$, $q$ and $\rho_{\text{sp}}$ can alter the horizon as expected. 
Furthermore, the presence of the BGGM suggests that our spacetime is not asymptotically flat.

\begin{figure}[htb]
	\centering 
	\includegraphics[width=.5\textwidth]{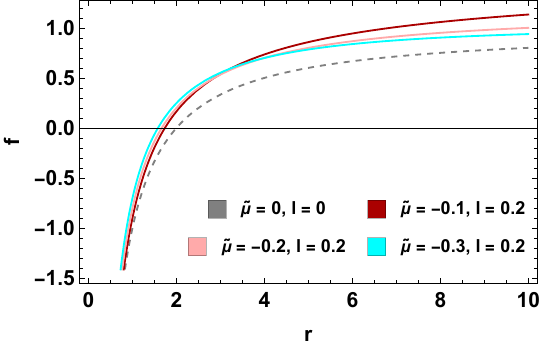}
	\caption{\label{fig:MetricFuncTX} 
		Plots of the metric function $\text{f}(r)$ as a function of the radial coordinate, showing the impact of varying $\tilde{\mu}$ for the $\text{M87}^{*}$ BH surrounded by a DM spike.
		The dashed gray lines indicate the scenario without the BGGM effect.}
\end{figure}

To show the behavior of the metric function $\text{f}(r)$, we plot Eq. \eqref{MetCoeffFunc-f11} in Figure~\ref{fig:MetricFuncTX}, with all parameters scaled to the BH parameters. It is observed that $\text{f}(r)$ reveals considerable variations for higher values of $\rho_{\text{sp}}$ and model parameters, both in the vicinity of the BH and at larger distances.

\section{Normalization of the DM spike profile
	\label{sec3}
}
Based on the observational data  for $\text{Sgr A}^{*}$ and $\text{M87}^{*}$ SMBHs taken from Refs. \cite{GorchteinPRD2010,LacroixPRD2015,LacroixPRD2017,NampalliwarAJ2021,DaghighApJ2022,PantigEPJC2022}, and following the procedure provided in Refs. \cite{GorchteinPRD2010,LacroixPRD2015,LacroixPRD2017}, we determine the normalization parameter $\rho_{0}$ for the DM spike profile \eqref{densityDMS1} at the centers of both the Milky Way and $\text{M87}$ galaxies, with the halo parameter $r_{0}$ being fixed at $20\, \text{kpc}$ for both galaxies.

We find the value of $\rho_{0}$ by ensuring that the density profile aligns with both the total mass of the galaxy and the mass encompassed within the radius of influence of the SMBHs, typically around $10^{5} R_{\text{s}}$. Thus, the DM mass within the relevant region for determining the BH mass, typically within $R_{0}=10^{5} R_{\text{s}}$, must be smaller than the uncertainty on the BH mass $\Delta M_{\text{BH}}$.  In order to determine $\rho_{0}$, we solve the following integral equation
\begin{equation}\label{NormInt}
	\int_{r_{\text{b}}}^{10^{5} R_{\text{s}}} 4\pi r^{2} \rho_{\text{DM}}(r) dr = \Delta M_{\text{BH}},
\end{equation}
in which we can suppose that the DM density profile vanishes below a certain radius, denoted as $r_{\text{b}}$, which is equal to either $4 R_{\text{s}}$ for the Newtonian approximation \cite{GondoloPRL1999} or $2 R_{\text{s}}$ for the full relativistic case of a static BH \cite{SadeghianPRD2013}. This occurs due to the capture of DM particles by the BH.
We encounter an additional complication due to the non-linear dependence of $\rho_{0}$ in the integral mentioned above, which arises from its correlation with $\rho_{\text{DM}}$. However, we can simplify this situation by considering that the mass is primarily dominated by contributions from regions where $r \gg R_{\text{s}}$, typically $r > r_{\text{b}} \rightarrow \mathcal{O}(100 R_{s})$. In this regime, we can approximate $\rho_{\text{DM}}\simeq \rho_{\text{DM}}^{\text{sp}}$. Moreover, we may factorize the dependence of $\rho_{0}$ in $\rho_{\text{DM}}^{\text{sp}}$ as $\rho_{\text{DM}}^{\text{sp}}(r) = \rho_{0}^{1/(4-\gamma)} ( \tilde{R}_{\text{sp}}/ r_{0})^{-\gamma}( \tilde{R}_{\text{sp}}/ r)^{\gamma_{\text{sp}}}$, where $\tilde{R}_{\text{sp}} = \mathcal{N}_{\gamma}r_{0}(M_{\text{BH}}/r_{0}^{3})^{1/(3-\gamma)}$ .  By doing so, Eq. \eqref{NormInt} becomes linear in $\rho_{0}$, leading us to the following expression
\begin{equation}\label{NormPara}
	\rho_{0} = \pi^{\gamma-4}\left(\frac{\left(\frac{\tilde{R}_{\text{sp}}}{r_{0}}\right)^{-\gamma}\left(4r_{\text{b}}^{3}\left(\frac{\tilde{R}_{\text{sp}}}{r_{0}}\right)^{\gamma_{\text{sp}}}-4R_{0}^{3}\left(\frac{\tilde{R}_{\text{sp}}}{R_{0}}\right)^{\gamma_{\text{sp}}}\right)}{\left(\gamma_{\text{sp}}-3\right)\Delta M_{\text{BH}}}
	\right)^{\gamma-4}.
\end{equation}
In this paper, we will adopt $\gamma = 1$ as the value for practical purposes, associated with the NFW density profile. For this choice, the associated DM spike profile possesses a power-law index of $\gamma_{\text{sp}} = 7/3$.

\subsection{The Milky Way galaxy}
Now, by using the observational data for the $\text{Sgr A}^{*}$ SMBH, located at the center of the Milky Way galaxy, with a mass of $M_{\text{BH}}=4.1\times10^{6}\,M_{\odot}$ and an uncertainty in mass of $\Delta{M_{\text{BH}}}=3\times10^{7}\,M_{\odot}$, along with the corresponding Schwarzschild radius $R_{\text{s}}\simeq 3.9\times 10^{-7}\,\text{pc}$, we obtain the data that appear in Table~\ref{table111}, that includes the value for $\mathcal{N}_{\gamma}$, the derived $\rho_{0}$, and the radius $R_{\text{sp}}$, the density $\rho_{\text{sp}}$, and the total mass of the DM spike $M_{\text{tot}}^{\text{sp}}$, obtained using Eqs. \eqref{densityDMS1}, \eqref{DMSpikeMass1} and \eqref{NormPara}.

\begin{table}[htb]
	\caption{The DM spike's parameters for the Milky Way galaxy.}
	\centering
	\setlength{\tabcolsep}{3pt} 
	\renewcommand{\arraystretch}{1.2} 
	\begin{tabular}{lccccr}
		\hline
		$\gamma_{\text{sp}}$ & $\mathcal{N}_{\gamma}$ & $\rho_{0} (\text{M}_\odot\,\text{pc}^{-3})$ & $R_{\text{sp}} (\text{pc})$ & $\rho_{\text{sp}} (\text{M}_\odot\,\text{pc}^{-3})$ & $M_{\text{tot}}^{\text{sp}} (\text{M}_\odot)$ \\
		\hline
		9/4 & 0.1 & $2.53\times10^{18}$ & $1.20 \times 10^{-5}$ & $2.53\times10^{18}$ & $6.38\times10^{4}$ \\
		\hline
		7/3 & 0.1 & $8.04\times 10^{7}$ & $1.60 \times 10^{-4}$ & $1.00 \times 10^{16}$ & $7.50\times10^{5}$ \\
		\hline
	\end{tabular}\label{table111}
\end{table}

\subsection{The $\text{M87}$ galaxy}
In a similar way, we consider the available data for the SMBH $\text{M87}^{*}$, which has a mass of $M_{\text{BH}}=6.4\times10^{9}\,M_{\odot}$ and an uncertainty in mass of $\Delta{M_{\text{BH}}}=5\times10^{8}\,M_{\odot}$. The corresponding Schwarzschild radius is approximately $R_{\text{s}}\simeq 6 \times 10^{-4}\,\text{pc}$. Based on this information, we obtain the data shown in Table~\ref{table222}, which includes the value for $\mathcal{N}_{\gamma}$, the derived $\rho_{0}$, and the radius $R_{\text{sp}}$, the density $\rho_{\text{sp}}$, and the total mass of the DM spike $M_{\text{tot}}^{\text{sp}}$. These quantities are obtained using Eqs. \eqref{densityDMS1}, \eqref{DMSpikeMass1}, and \eqref{NormPara}.

\begin{table}[htb]
	\caption{The DM spike's parameters for the $\text{M87}$ galaxy.}
	\centering
	\setlength{\tabcolsep}{3pt} 
	\renewcommand{\arraystretch}{1.2} 
	\begin{tabular}{lccccr}
		\hline
		$\gamma_{\text{sp}}$ & $\mathcal{N}_{\gamma}$ & $\rho_{0} (\text{M}_\odot\,\text{pc}^{-3})$ & $R_{\text{sp}} (\text{pc})$ & $\rho_{\text{sp}} (\text{M}_\odot\,\text{pc}^{-3})$ & $M_{\text{tot}}^{\text{sp}} (\text{M}_\odot)$ \\
		\hline
		9/4 & 0.1 & $1.43\times10^{4}$ & $7.65$ & $1.43\times10^{4}$ & $1.07\times10^{8}$ \\
		\hline
		7/3 & 0.1 & $6.52 \times10^{-2}$ & $221.54$  & $5.89$ &  $1.21\times10^{9}$\\
		\hline
	\end{tabular}\label{table222}
\end{table}

\section{The Kerr bumblebee BH with a global monopole in DM spike
	\label{sec4}
}
In Section 2, we derived the BGMSLBH solution, which is influenced by the bumblebee field in the presence of the GM, along with the surrounding DM spike. We now begin with a general seed spherically symmetric and static metric expressed as follows:
\begin{equation}\label{MetricBBH1}
	\mathrm{d}s^{2}=-\mathcal{F}(r)dt^{2}+\frac{1}{\mathcal{G}(r)}dr^{2}+\mathrm{h}(r)\left(d\theta^{2}+\sin^{2}\theta d\varphi^{2}\right).
\end{equation}
With the idea of extending our BGMSLBH solution to incorporate rotation, introducing a spin parameter denoted as $a$, using the modified NJ method \cite{AzregPRD2014,AzregEPJC2014,AzregPLB2014,JohannsenPRD2011,ToshmatovPRD2014,ToshmatovEPJP2017} following the approach presented in \cite{AzregPRD2014,AzregEPJC2014,AzregPLB2014}.
The modified version of the NJ method  that we employ here differs from the original one \cite{NewmanJMP1965} by excluding one of its steps, specifically, the complexification of coordinates. Instead, we utilize an alternate coordinate transformation as given in \cite{AzregPRD2014}. 
Moreover, this modified NJ method has been applied in several significant studies within the framework of rotating DM-BH systems \cite{XuJCAP2021,JusufiPRD2019,JusufiMNRA2021,PantigJCAP2022, Arturo,TangJHEP2022,XuPRD2017}.  By applying the modified NJ method, we derive the metric describing the Kerr Bumblebee BH spacetime with GM, encompassed by the DM spike as follows
\begin{subequations}\label{RotatingSTmetric2}
	\begin{equation}
			ds^{2} = -\left(1-\frac{\Psi}{\Sigma^{2}}\right)d{t}^{2}+\frac{\Sigma^{2}}{\Delta}\,d{r}^{2}-2a\, \frac{\Psi}{\Sigma^{2}}\, \text{sin}^{2}\theta \,d{t}d{\varphi}
			 +\Sigma^{2}\,d{\theta}^{2}+\frac{\Xi\, \text{sin}^{2}\theta}{\Sigma^{2}}d{\varphi}^{2},
	\end{equation}
	defining new notations
	\begin{align}\label{SigPsiXiDM}
		\nonumber
		&\Sigma^{2}  = \mathrm{K}+a^{2}\text{cos}^{2}\theta,\qquad\qquad\quad 
		\Psi = \mathrm{K} - \mathcal{G}(r)\mathrm{h}(r),\\
		& \mathrm{K}(r)=\mathrm{h}(r) \sqrt{\frac{\mathcal{G}(r)}{\mathcal{F}(r)}},\qquad \qquad\
		\Delta(r)=\mathcal{G}(r)\mathrm{h}(r)+a^{2},\\
		\nonumber
		& \Xi = \left(\mathrm{K}+a^2\right)^{2}-a^2 \Delta\, \text{sin}^{2}\theta.
	\end{align}
\end{subequations}
The modified Kerr BH metric \eqref{RotatingSTmetric2} returns to the Kerr BH metric as the BGGM parameters $w$ and $q$ approaches one, assuming the presence of the DM spike. 
Similarly, in the absence of the DM spike, the metric reverts to the standard Kerr BH metric as the limit $\rho_{\text{sp}} \rightarrow 0$ (or $r \rightarrow r_{\text{b}}$) is taken. Additionally, when the parameter $a$ tends to zero, the modified Kerr BH spacetime converges to the spherically symmetric metric \eqref{MetricBBH1}.
Similar to the Kerr spacetime, the modified Kerr BH spacetime also possesses two Killing vectors denoted by $\xi^{\mu}_{(t)}=\delta^{\mu}_{t}$ and $\xi^{\mu}_{(\varphi)}=\delta^{\mu}_{\varphi}$, which remain invariant under transformations involving the time coordinate $t$ and the azimuthal angle $\varphi$.

Let us now investigate the effects of the BGGM and the DM distribution within the spike region, described by the density profile in Eq. \eqref{densityDMS1}, on the horizon, SLS, and ergoregion \footnote{ The region lying between the SLS and the event horizon can be defined using $\delta_{\text{er}}=r_{\text{SLS}}^{+}-r_{+}$.} of the BGMKLBH. Due to the complexity arising from the DM mass function appearing in solving $\Delta = 0$ and $g_{tt} = 0$, we need to numerically analyze the impacts of the spike profile and the BGGM theoretically through some plots and tables. Subsequently, we will apply this analysis to the cases of the $\text{M87}$ and Milky Way galaxies, which host the well-known SMBHs $\text{M87}^{*}$ and $\text{Sgr A}^{*}$, respectively. To do this, we will make use of the observational data for the mentioned SMBHs and their corresponding galaxies, as presented in Section 3, for the DM spike profile. 
It is worth noting that the observational data for the DM spike profile in the Milky Way galaxy, in terms of the parameters of the SMBH $\text{Sgr A}^{*}$, can be expressed as $R_{\text{sp}}\simeq407.16R_{\text{s}}$ and $\rho_{\text{sp}}\simeq6.15 \times 10^{-10}\rho_{\text{BH}}$, where $\rho_{\text{BH}} = M_{\text{BH}}/(4\pi R_{\text{s}}/3)$. Similarly, the observational data for the spike profile in the $\text{M87}$  galaxy, associated with the $\text{M87}^{*}$ SMBH parameters, can be represented as $R_{\text{sp}}\simeq3.7 \times 10^{5}R_{\text{s}}$ and $\rho_{\text{sp}}\simeq 8.32 \times 10^{-19}\rho_{\text{BH}}$.

\begin{figure}[htb]
	\centering 
	\includegraphics[width=.45\textwidth]{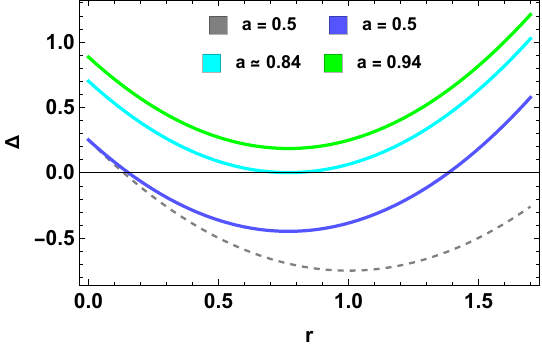}
		\hfill
	\includegraphics[width=.45\textwidth]{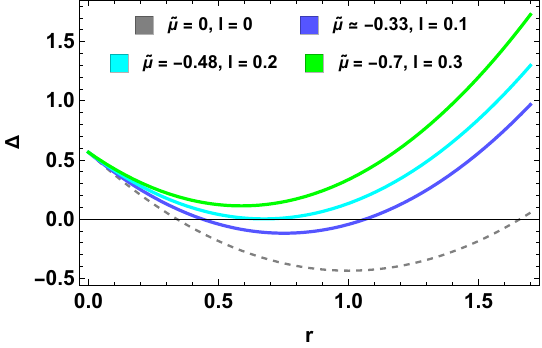}
	\caption{\label{fig:Sec31}
		Plot illustrating the horizon behavior $(\Delta\, \text{vs}\, r)$ in the BGMKLBH spacetime for the $\text{M87}$ galaxy. The left panel shows results for fixed values of $\tilde{\mu} = -0.3$ and $\ell = 0.1$ with varying $a$, while the right panel depicts fixed $a = 0.75$ with varying $\tilde{\mu}$ and $\ell$. We have set $M_{\text{BH}} = 1$ in all cases. The dashed gray lines indicate the scenario without the BGGM effect.}
\end{figure}

\begin{figure*}[htb]
	\centering 
	\includegraphics[width=.45\textwidth]{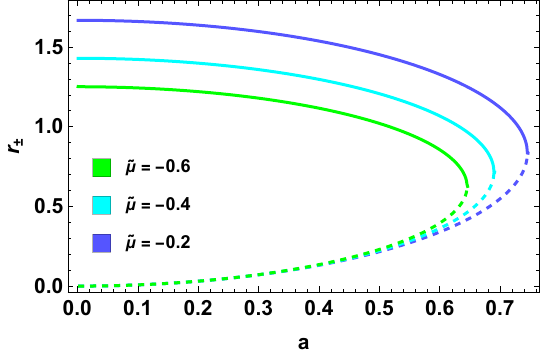}
	\hfill
	\includegraphics[width=.45\textwidth]{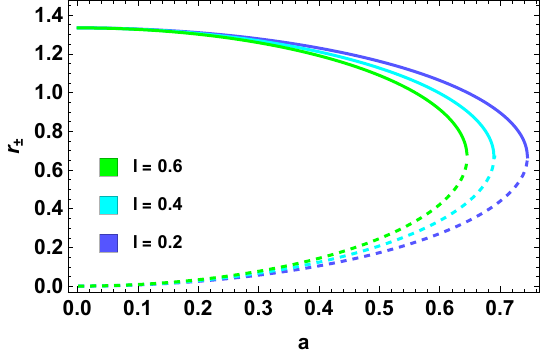}
	\\
	\includegraphics[width=.45\textwidth]{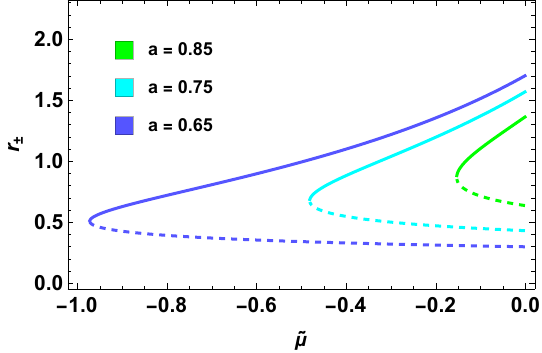}
	\hfill
	\includegraphics[width=.45\textwidth]{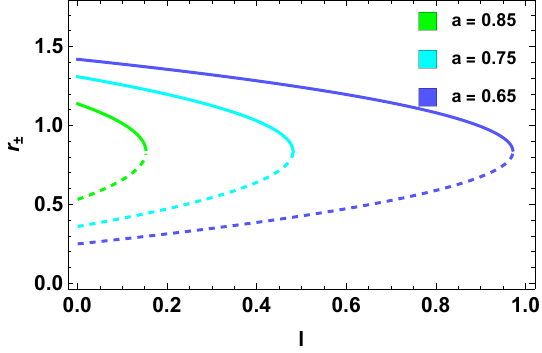}
	\caption{\label{fig:Sec32}
		Plot illustrates the variation of the event horizon $r_{+}$ (solid lines) and the Cauchy horizon $r_{-}$ (dashed lines) concerning the spin parameter $a$ and model parameters $\ell$ and $\tilde{\mu}$ for the BGMKLBH.
	}
\end{figure*}

We observe that the influence of DM on the horizon and SLS, and subsequently on the ergoregion, primarily depends on the mass distribution in close proximity to the SMBH at the center of the galaxy. Therefore, we can neglect the contribution of DM located far from the spike, particularly in regions where $r$ significantly exceeds $R_{\text{sp}}$.
Now, our focus shifts towards understanding the characteristics of the BGMKLBHs given in \eqref{RotatingSTmetric2}. The objective here is to demonstrate that their attributes closely resemble those of the Kerr BH. Specifically, we explore the horizons and SLSs and delineate the region situated between the corresponding event horizon and the outer SLS. 
This region is known as the ergoregion \cite{GhoshJCAP2021,KumaraJCAP2020}. We intend to analyze the effects of BGGM on the ergoregion structure and its two boundaries -- namely, the event horizon and the outer SLS. 
The horizons of the BGMKLBH can be considered as the solutions to the following equation
\begin{equation}\label{Horzions1}
	\mathcal{G}(r)\mathrm{h}(r)+a^{2} = 0,
\end{equation}
which also corresponds to the coordinate singularity of the metric \eqref{RotatingSTmetric2}.  Through numerical analysis, it becomes apparent that, contingent upon the values of the parameters in the DM spike mass function, the spin parameter $a$, and the BGGM parameters $\tilde{\mu}$ and $\ell$, Eq. \eqref{Horzions1} can yield a maximum of two distinct real positive roots, degenerate roots, or no-real positive roots. These outcomes respectively correspond to non-extremal BH configurations, extremal BH configurations, and no-BH configurations for metric \eqref{RotatingSTmetric2}. The two real positive roots of Eq. \eqref{Horzions1} are recognized as the radii of the Cauchy horizon $(r_{-})$ and the event horizon $(r_{+})$, with the condition that $r_{-} \leq r_{+}$ (as shown in Figures~\ref{fig:Sec31} and \ref{fig:Sec32}).
Figures~\ref{fig:Sec31} and \ref{fig:Sec32}, as well as Tables~\ref{Tab:Sec31} and \ref{Tab:Sec32}, illustrate the behavior of the horizon radii $r_{\pm}$ as the spin parameter $a$ and model parameters vary. 

For given values of DM spike parameters, as the spin parameter $a$ and the model parameters $\ell$ and $\tilde{\mu}$ vary, the event horizon radius $r_{+}$ decreases (increases), while the Cauchy horizon radius $r_{-}$ increases (decreases) with rising $a$ and $\ell$ ($\tilde{\mu}$)
\footnote{Please refer to Figure~\ref{fig:Sec32}, as well as Tables~\ref{Tab:Sec31} and \ref{Tab:Sec32}.}. 
For a fixed value of $a$ ($\ell$ and $\tilde{\mu}$), there exists a critical value $\ell_{\text{ex}}$ and $\tilde{\mu}_{\text{ex}}$ ($a_{\text{ex}}$) where both horizons coincide, resulting in $r_{-} = r_{+}$. 
This signifies that for $\tilde{\mu} < \tilde{\mu}_{\text{ex}}$ and $\ell > \ell_{\text{ex}}$ ($a > a_{\text{ex}}$), Eq. \eqref{Horzions1}  possesses no roots, and for $\tilde{\mu} > \tilde{\mu}_{ex}$ and $\ell < \ell_{\text{ex}}$ ($a < a_{ex}$), two distinct roots emerge (as can be seen in Figure~\ref{fig:Sec31}).
Furthermore, for consistent values of the DM spike parameter and spin $a$, the presence of the BGGM parameters leads to a reduction in $r_{+}$ and an increase in $r_{-}$. 
Moreover, the GM effect contributes more significantly to this scenario than the BG effect.

\begin{figure*}[htb]
	\centering 
	\includegraphics[width=.24\textwidth]{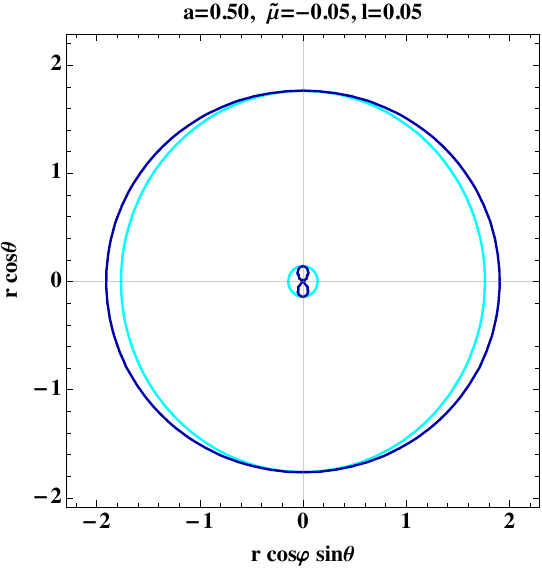}
	\hfill
	\includegraphics[width=.24\textwidth]{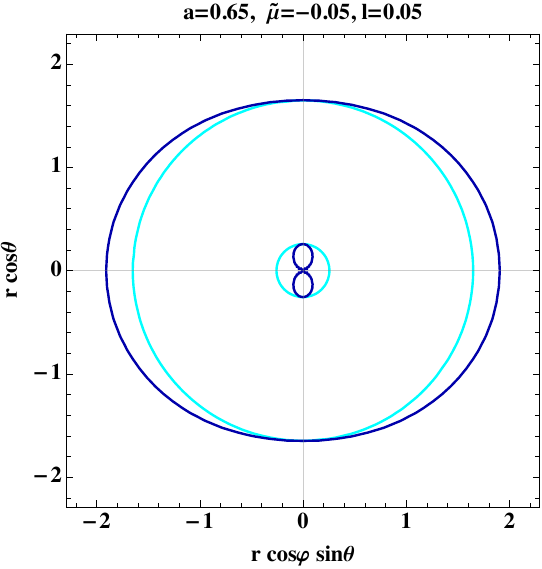}
	\hfill
	\includegraphics[width=.24\textwidth]{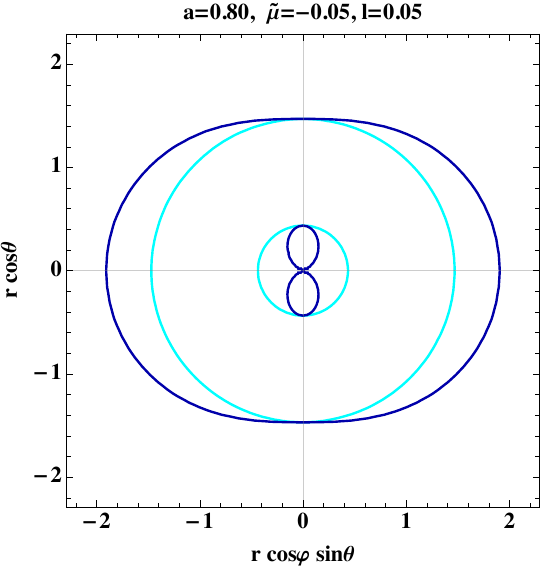}
	\hfill
	\includegraphics[width=.24\textwidth]{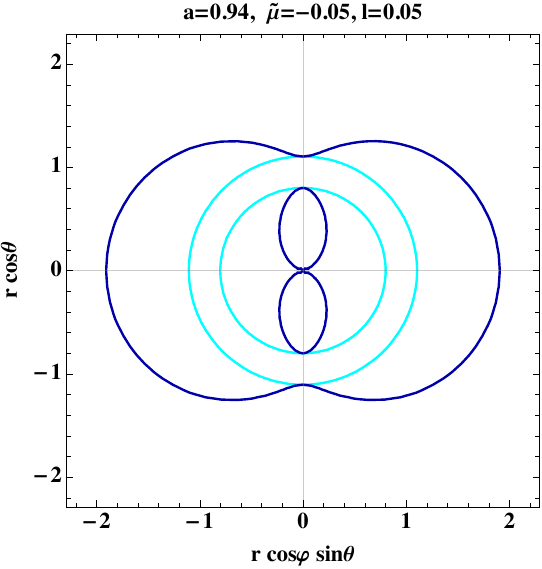}
	\hfill
	\includegraphics[width=.24\textwidth]{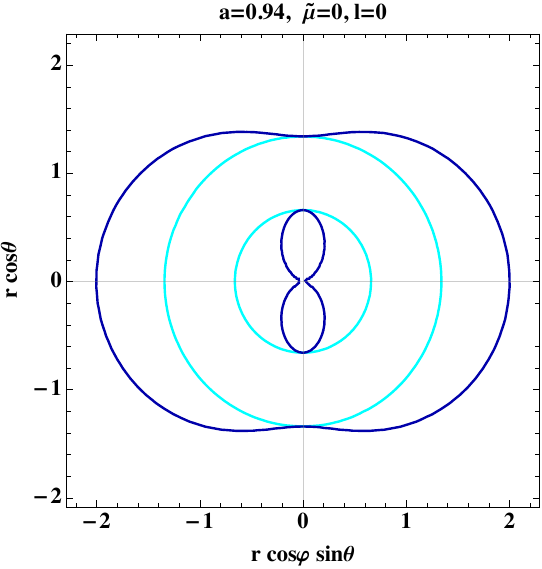}
	\includegraphics[width=.24\textwidth]{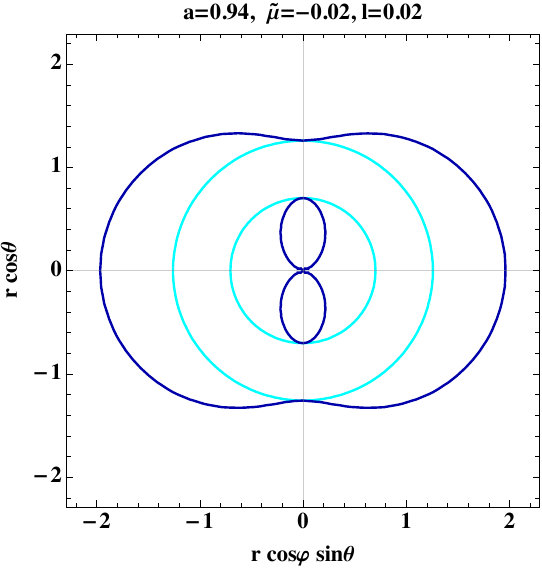}
	\hfill
	\includegraphics[width=.24\textwidth]{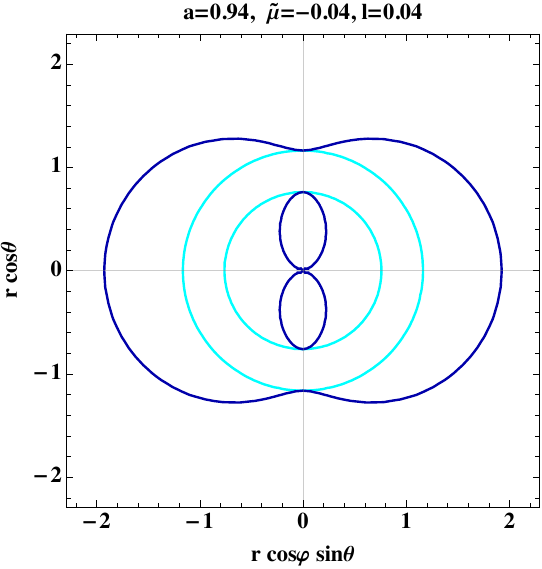}
	\hfill
	\includegraphics[width=.24\textwidth]{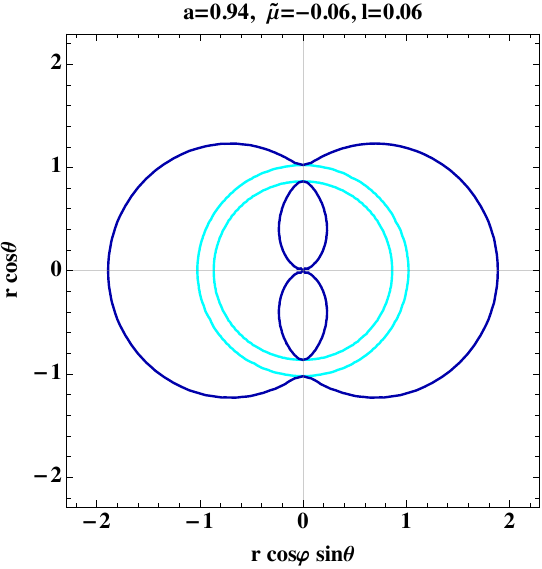}
	\caption{\label{fig:Sec33} 
		Visualizing the behavior of the ergoregion in the xz-plane of the BGMKLBH spacetime  corresponding to the $\text{M87}^{*}$ BH with the dark blue and cyan lines assigned to the SLS and horizons, respectively. 
	}
\end{figure*}

In the absence of BH spin, the SLS overlap with the BH's event horizon. However, in the presence of non-zero spin, two SLSs emerge positioned distinctively apart from the event horizon, viz., the outer SLS ($r_{\text{SLS}}^{+}$) and inner SLS ($r_{\text{SLS}}^{-}$) \cite{XuJCAP2021}. This scenario holds true for the BGMKLBH metric given in Eq. \eqref{RotatingSTmetric2} as well.
Now, let us proceed to determine the SLS for the BGMKLBH in the $\mathcal{F} \neq \mathcal{G}$ spacetime configuration. This can be achieved by considering the condition $g_{tt}=0$, which can be expressed as
\begin{equation}
	\Sigma^{2}-\Psi=0,
\end{equation}	
where, $\Sigma^{2}$ and $\Psi$ are defined in Eq. \eqref{SigPsiXiDM}.
Through numerical computation of this equation and its integration with the event horizon findings of the BGMKLBH, it becomes evident that, aside from the parameters related to BH and the DM spike masses, changes occur in  $r_{+}$, $r_{\text{SLS}}^{+}$ and $\delta_{\text{er}}$ as $a$, $\tilde{\mu}$ and $\ell$ vary. 
Specifically, when $\ell$ increases and $\tilde{\mu}$ decreases individually while keeping $a$ constant, $r_{\text{SLS}}^{+}$ decreases and $\delta_{\text{er}}$ increases. Conversely, for fixed values of $\ell$ and $\tilde{\mu}$, an increase in $a$ leads to an decrease in $r_{\text{SLS}}^{+}$ and a increase in $\delta_{\text{er}}$
(cf. Figures~\ref{fig:Sec33} as well as Tables~\ref{Tab:Sec31}, and \ref{Tab:Sec32}).

\begin{table}[htb]
	\centering
	\caption{The horizons, SLSs, and the region between the SLS and the event horizon of the BGMKLBH spacetime related to the $\text{M87}$ galaxy, showcasing variations based on different values of $a$, $\tilde{\mu}$ and $\ell$. \\
		\label{Tab:Sec31}	
	} 
	\begin{tabular}{lcccr} 
		\hline
		$\tilde{\mu}$ & $r_{-}$ & $r_{+}$ & $r_{\rm SLS}^{+}$&$\delta_{\rm er}$\\
		\hline
		&&(I) $a=0.75$, $\ell = 0.01$&&\\[0.5mm]
		\,\,\,0.00 & 0.34285 & 1.65711 & 1.84608 & 0.18897 \\
		- 0.01  & 0.34375 & 1.63640 & 1.82614 & 0.18974 \\
		- 0.02  & 0.34466 & 1.61608 & 1.80658 & 0.19050 \\
		- 0.03  & 0.34558 & 1.59612 & 1.78740 & 0.19128 \\
		\hline
		&&(II) $a=0.5$, $\ell = 0.01$&&\\[0.5mm]
		\,\,\,0.00 & 0.13543 & 1.86453 & 1.93470 & 0.07017 \\
		- 0.01  & 0.13553 & 1.84462 & 1.91487 & 0.07025 \\
		- 0.02  & 0.13564 & 1.82510 & 1.89544 & 0.07034 \\
		- 0.03  & 0.13575 & 1.80596 & 1.87638 & 0.07042 \\
		\hline
		$\ell$ & $r_{-}$ & $r_{+}$ & $r_{\rm SLS}^{+}$&$\delta_{\rm er}$\\
		\hline
		&&(I) $a=0.75$, $\tilde{\mu} = -0.01$&&\\[0.5mm]
		0.00 & 0.33946 & 1.64070 & 1.82780 & 0.18710 \\
		0.01  & 0.34375 & 1.63640 & 1.82614 & 0.18974 \\
		0.02  & 0.34808 & 1.63208 & 1.82447 & 0.19239 \\
		0.03  & 0.35243 & 1.62773 & 1.82280 & 0.19507 \\
		\hline
		&&(I) $a=0.5$, $\tilde{\mu} = -0.01$&&\\[0.5mm]
		0.00 & 0.13409 & 1.84607 & 1.91554 & 0.06947 \\
		0.01  & 0.13553 & 1.84462 & 1.91487 & 0.07025 \\
		0.02  & 0.13698 & 1.84317 & 1.91421 & 0.07104 \\
		0.03  & 0.13844 & 1.84172 & 1.91354 & 0.07182 \\
		\hline
	\end{tabular}
\end{table}

\begin{table}[htb]
	\centering
	\caption{The horizons, SLSs, and the region between the SLS and the event horizon of the BGMKLBH spacetime related to the Milky Way galaxy, showcasing variations based on different values of $a$, $\tilde{\mu}$ and $\ell$.\\
		\label{Tab:Sec32}
	} 
	\begin{tabular}{lcccr} 
		\hline
		$\tilde{\mu}$ & $r_{-}$ & $r_{+}$ & $r_{\rm SLS}^{+}$&$\delta_{\rm er}$\\
		\hline
		&&(I) $a=0.75$, $\ell = 0.01$&&\\[0.5mm]
		\,\,\,0.00 & 0.34511 & 1.65110 & 1.84137 & 0.19027 \\
		- 0.01  & 0.34603 & 1.63037 & 1.82142 & 0.19105 \\
		- 0.02  & 0.34696 & 1.61002 & 1.80186 & 0.19184 \\
		- 0.03  & 0.34791 & 1.59004 & 1.78268 & 0.19264 \\
		\hline
		&&(II) $a=0.5$, $\ell = 0.01$&&\\[0.5mm]
		\,\,\,0.00 & 0.13618 & 1.85993 & 1.93045 & 0.07052 \\
		- 0.01  & 0.13629 & 1.84001 & 1.91062 & 0.07061 \\
		- 0.02  & 0.13640 & 1.82048 & 1.89118 & 0.07070 \\
		- 0.03  & 0.13651 & 1.80133 & 1.87211 & 0.07078 \\
		\hline
		$\ell$ & $r_{-}$ & $r_{+}$ & $r_{\rm SLS}^{+}$&$\delta_{\rm er}$\\
		\hline
		&&(I) $a=0.75$, $\tilde{\mu} = -0.01$&&\\[0.5mm]
		0.00 & 0.34170 & 1.63470 & 1.82310 & 0.18840 \\
		0.01  & 0.34603 & 1.63037 & 1.82142 & 0.19105 \\
		0.02  & 0.35040 & 1.62601 & 1.81975 & 0.19374 \\
		0.03  & 0.35479 & 1.62162 & 1.81807 & 0.19645 \\
		\hline
		&&(I) $a=0.5$, $\tilde{\mu} = -0.01$&&\\[0.5mm]
		0.00 & 0.13483 & 1.84146 & 1.91129 & 0.06983 \\
		0.01  & 0.13629 & 1.84001 & 1.91062 & 0.07061 \\
		0.02  & 0.13775 & 1.83855 & 1.90994 & 0.07139 \\
		0.03  & 0.13921 & 1.83709 & 1.90927 & 0.07218 \\
		\hline
	\end{tabular}
\end{table}

\section{The shadow of a deformed Kerr BH in DM spike
	\label{sec5}}
The BH shadow emerges as the optical manifestation resulting from the strong GL of light in the vicinity of these compact entities, presenting itself as a two-dimensional dark region when observed from a considerable distance. 
Now, 
we consider BGMKLBH as SMBH located at the center of the $\text{M87}$ galaxy. 
In this section, we employ observational data of the SMBH and the DM spike profile within the $\text{M87}$ galaxy.
To explore the influence of the BGGM effect on the dark shadow cast by the BH within the DM spike, we investigate the impact of the parameter space $(a,\ell,\tilde{\mu})$ on the geodesic motion of photons around the BGMKLBH.  Additionally, the size and shape of the shadows are found to be dependent on the BGMKLBH parameters.
The shadow, in turn, has played a crucial role in estimating and measuring various BH parameters, such as mass and spin angular momentum, as well as facilitating the estimation of the BGGM parameter associated with an extended gravity model \cite{CapozzielloRP2011,CaiRPP2016,NojiriPR2017,JusufiEPJC2022,LambiaseJCAP2023,LambiaseApJ2020,CapozzielloJCAP2017,CapozzielloPRD2014,CapozzielloAP2012,JohannsenPRL2016,HarkoPRD2009,PsaltisApJ2015,PsaltisLRR2008,KramerNAR2004}.

Our focus now turns to the study of photon geodesic motion within metric spacetime \eqref{RotatingSTmetric2}. 
We begin with the Lagrangian $\mathcal{L}$ for geodesic motion, given by
$2\mathcal{L}=g_{\mu\nu}\dot{x}^{\mu}\dot{x}^{\nu}$, where the values of $2\mathcal{L}$ are $-1, 0, 1$ respectively for timelike, null, and spacelike geodesics. Here, the overdot denotes a derivative concerning the affine parameter $\lambda$, connected to proper time via $\tau=\lambda m$, wherein $m$ signifies the mass of the test particle. However, in the case of photons, we have $m=0$.
In this context, the null geodesics can be derived by employing the method of separating variables within the Hamilton-Jacobi equation \cite{CarterPR1968,JohannsenPRD2013,HiokiPRD2008}
\begin{equation}\label{HamiltonJacobiEq}
	\mathcal{H} = -\frac{\partial S}{\partial \lambda} = \frac{1}{2} g_{\mu\nu} \frac{\partial S}{\partial x^{\mu}} \frac{\partial S}{\partial x^{\nu}} =0,
\end{equation}
in which $\mathcal{H}$ and $S$ represent the canonical Hamiltonian and the Jacobi action, respectively.
Given that the BGMKLBH spacetime \eqref{RotatingSTmetric2} remains independent of $t$ and $\varphi$, these coordinates can be regarded as cyclic, and the associated Killing vectors are expressed as $\xi^{\mu}_{(t)}=\delta^{\mu}_{t}$ and $\xi^{\mu}_{(\varphi)}=\delta^{\mu}_{\varphi}$. Hence, we are able to introduce the conserved energy $E$ and the $z$ component of angular momentum $L_{z}$, being their explicit definitions 
\begin{equation}
	E= -\frac{\partial S}{\partial t} = - g_{tt}\dot{t} - g_{\varphi t}\dot{\varphi}, \quad \text{and} \quad L_{z} = \frac{\partial S}{\partial \varphi}= g_{\varphi t}\dot{t} + g_{\varphi \varphi}\dot{\varphi}.
\end{equation}
To obtain a separable solution for Eq. \eqref{HamiltonJacobiEq}, we can represent the Jacobi action in terms of the constants of motion $E$ and $L_{z}$ as follows:
\begin{equation}\label{JacobiAction}
	S=-Et+L_{z}\varphi+S_{r}(r)+S_{\theta}(\theta),
\end{equation}
where $S_{r}(r)$ and $S_{\theta}(\theta)$ exclusively depend on the $r$ and $\theta$ coordinates, respectively.
In the geodesic equations of motion, there exist four constants of motion that enable us to formulate the null geodesic equation in a first-order format: the Lagrangian $\mathcal{L} = 0$, the energy $E$, the $z$ component of angular momentum $L_{z}$, and the Carter constant $\mathcal{K}$. This final constant of motion can appear through the separation process of the Hamilton-Jacobi equation. By inserting the Jacobi action \eqref{JacobiAction} into Eq. \eqref{HamiltonJacobiEq}, the resulting expression becomes:
\begin{equation}\label{SeparatedEq}
	-\Delta \left(\frac{\partial S_{r}}{\partial r}\right)^{2}+\frac{\left(\mathcal{U}(r)E-aL_{z}\right)^{2}}{\Delta}=\left(\frac{\partial S_{\theta}}{\partial \theta}\right)^{2}+\frac{\left(L_{z}-aE \sin^{2}\theta\right)^{2}}{\sin^{2}\theta}=\mathcal{K},
\end{equation}
where each side depends on either $r$ or $\theta$. This implies that both sides are equal the Carter constant $\mathcal{K}$. Besides, $\mathcal{U}(r) = \mathrm{h}(r)\sqrt{\mathcal{G}(r)/\mathcal{F}(r)}+a^{2}\equiv \mathrm{K}(r)+a^{2}$. 
In the absence of the BGGM effect, where $q \rightarrow 1$ and $w \rightarrow 1$ (or $\ell \rightarrow 0$ and $\tilde{\mu} \rightarrow 0$), and without the DM spike, as $\rho_{\text{sp}} \rightarrow 0$ (or $r \rightarrow r_{\text{b}}$), the function $\mathcal{U}(r)$ reduces to that of the standard Kerr BH metric \cite{KerrPRL1963}, which is $(r^{2}+a^{2})$.
Following this, we derive four first-order differential equations describing the geodesic motions in the vicinity of the BGMKLBH, which is influenced by a distribution of DM and the BGGM effect. Thus, the equations of motion are as follows
\begin{subequations}
	\begin{align}
		& \Sigma^{2} \dot{t} = a \left(L_{z} - a E\, \text{sin}^{2}\theta\right)+\frac{\mathcal{U}}{\Delta}\left(\mathcal{U}E-a L_{z}\right), \label{geoEq1}\\
		&\Sigma^{2} \dot{\varphi} =  \frac{L_{z}}{\text{sin}^{2}\theta} -a E  +\frac{a}{\Delta} \left(\mathcal{U}E -a L_{z}\right) ,\label{geoEq2}\\
		&\Sigma^{2} \dot{r} =\pm \sqrt{R(r)},\label{geoEq3}\\
		&\Sigma^{2} \dot{\theta} =\pm \sqrt{\Theta(\theta)},\label{geoEq4}
	\end{align}
\end{subequations}
where
\begin{subequations}
	\begin{align}
		&R(r) \equiv \Delta^{2}\left(\frac{\partial S_{r}}{\partial r}\right)^{2}=E^{2} \left[\left(\mathcal{U} -a \xi\right)^{2}-\Delta\,\mathcal{C}\right],\label{FunctionR}\\
		&\Theta(\theta) \equiv \left(\frac{\partial S_{\theta}}{\partial \theta}\right)^{2}=E^{2} \left[\mathcal{C}-\left(a \sin\theta-\frac{\xi}{\sin\theta}\right)^{2}\right], \label{Functiontheta}
	\end{align}
\end{subequations}
and the dimensionless impact parameters are symbolized as  $\xi=L_{z}/E$ and $\eta=\mathcal{K}/E^{2}$.
The impact parameters $\xi$ and $\eta$ have a connection with the constant $\mathcal{C}$ given by $\mathcal{C} =\eta+ \left(a-\xi\right)^{2}$ \cite{HiokiPRD2009,JusufiEPJC2020354,Chandrasekhar,JohannsenAPJ2013,TsukamotoJCAP2014,TsukamotoPRD2018,KocherlakotaPRD2021,StaelensPRD2023,TeoGRG2021,GottCQG2019,KumarApJ2020-1,KumarApJ2020-2}. 
In examining the BH shadow, our focus lies on the unstable circular photon orbits. This involves satisfying conditions: $R(r_{\text{ph}})=0$, $R'(r_{\text{ph}})=0$, and $R'' \leq 0$, with $r=r_{\text{ph}}$ denoting the unstable photon orbit radius.
Using the aforementioned conditions, the critical impact parameters $(\xi_c, \eta_c)$ for the unstable orbits that could determine the shape of the BH shadow can be obtained as 
\begin{subequations}\label{impactParameters}
	\begin{align}
		\xi_{c} & = \left. \frac{\mathcal{U}\Delta_{,r}-2\Delta\,\mathcal{U}_{,r}}{a \Delta_{,r}}\right|_{r\rightarrow r_{\mathrm{ph}}},\label{impactParameterXi}\\
		\eta_{c} & =\left. \frac{4 a^{2} \,\mathcal{U}^{2}_{,r}\Delta-\left(\left(\,\mathcal{U}-a^{2}\right)\Delta_{,r}-2\,\mathcal{U}_{,r} \Delta\right)^{2}}{a^{2}\Delta^{2}_{,r}}\right|_{r\rightarrow r_{\mathrm{ph}}}.\label{impactParameterEta}
	\end{align}
\end{subequations}

To explore how the BGGM model influences the shadow images of the BGMKLBH, we consider an observer positioned at coordinates $(r_{o}, \theta_{o})$, where $r_{o}$ represents the observer's distance and $\theta_{o}$ is the angular position in the sky. The BGMKLBH shadow shape relies on the deviation parameters, spin $a$, and the observation angle $\theta_{o}$ relative to the spin axis. By using the tetrad components of the four-momentum $p^{(\mu)}$, the connection between the observer?s celestial coordinates $(X,Y)$ and the critical impact parameters is given by:
\begin{subequations}\label{CelestialCoord1}
	\begin{equation}
		\begin{split}
			X &= -r_{0} \frac{p^{(\varphi)}}{p^{(t)}} = \left.-r_{0} \frac{\xi_{c}}{\sqrt{g_{\varphi\varphi}}\tilde{\zeta}\left(1+\frac{g_{t\varphi}}{g_{\varphi\varphi}}\xi_{c}\right)}\right|_{(r\rightarrow r_{0}, \, \theta\rightarrow\theta_{0})}, \\
			Y &= -r_{0} \frac{p^{(\theta)}}{p^{(t)}} = \left.\pm r_{0} \frac{\sqrt{ \eta_{c} + a^{2} \text{cos}^{2}\theta - \xi_{c}^{2}\text{cot}^{2}\theta}}{\sqrt{g_{\theta\theta}}\tilde{\zeta}\left(1+\frac{g_{t\varphi}}{g_{\varphi\varphi}}\xi_{c}\right)}\right|_{(r\rightarrow r_{0}, \, \theta\rightarrow\theta_{0})},
		\end{split}
	\end{equation}
	with 
	\begin{equation}
		\tilde{\zeta} = \sqrt{\frac{g_{\varphi\varphi}}{g^{2}_{t\varphi}-g_{tt}g_{\varphi\varphi}}}.
	\end{equation}
\end{subequations} 
In the case of non-asymptotically flat spacetime, which arises due to the existence of the BGGM background encompassing the combined influences of the GM and the spontaneous breaking of LS, 
and assuming that the observer is positioned at a finite distance away from the BH but still far away, with distances between the observer and the $\text{Sgr A}^{*}$ and $\text{M87}^{*}$ SMBHs being approximately $8.3 \,\text{kpc}$ and $16.8 \, \text{Mpc}$, respectively, 
the celestial coordinates given in Eq. \eqref{CelestialCoord1} can be expressed in a simplified form
\cite{JusufiPRD2019,JusufiEPJC2020354,NampalliwarAJ2021,CapozzielloJCAP2023,PantigJCAP2022}
\begin{equation}\label{CelestialCoord2}
	X  = -\sqrt{\text{f}(r_{0})} \frac{\xi_{c}}{\sin{\theta_{0}}}, \qquad Y=\pm \sqrt{\text{f}(r_{0})} \sqrt{ \eta_{c} + a^{2} \text{cos}^{2}\theta_{0} - \xi_{c}^{2}\text{cot}^{2}\theta_{0}}.
\end{equation}
For an observer situated in the equatorial plane with a latitude angle of $(\theta_{0}=\pi/2)$, Eq. \eqref{CelestialCoord2} can be reduced as
\begin{equation}\label{CelestialCoord3}
	X  = -\sqrt{\text{f}(r_{0})}\xi_{c}, \qquad  Y=\pm \sqrt{\text{f}(r_{0})} \sqrt{\eta_{c}}.
\end{equation}
Here, to delineate the BGMKLBH shadow, one can then plot $Y$ versus $X$, where the celestial coordinates $X$ and $Y$ satisfy the following relationship:
\begin{equation}\label{ShadowRadius1}
		X^{2}+Y^{2} = \text{f}(r_{0})\left(\xi_{c}^{2}+\eta_{c}\right)
		= \text{f}(r_{0}) \left(-a^{2}+2\, \mathcal{U}+\frac{4 \Delta\, \mathcal{U}_{,r}\left(\mathcal{U}_{,r}-\Delta_{,r}\right)}{\Delta_{,r}^{2}}\right),
\end{equation}
when the shadow is observed from the equatorial plane ($\theta_{0} = \pi/2$).
It is noteworthy to mention that the manifestation of a BH shadow, as perceived by an observer located at an infinite distance, is the result of the combined influence of all photon trajectories that do not intersect the photon sphere. Additionally, it is assumed that there is no internal light source within the photon sphere that could illuminate the shadow. The arrangement of light sources just influences the luminosity in the area surrounding the shadow, known as the photon shell, and does not influence the geometry of the shadow itself. Nevertheless, the geometric form is contingent upon the inclination angle $\theta_{0}$ \cite{FengEPJC2020}. Likewise, the greatest deformation of a BH shadow shape appears at its highest acceptable angular momentum when observed from a particular angle $\theta_0$. A DMS-BH interacting system with a particular mass, BGGM parameters, and angular momentum exhibits the highest distortion at $\theta_0=\pi/2$, which is also observable from the equatorial plane.

\begin{figure*}[htb]
	\centering 
	\includegraphics[width=.24\textwidth]{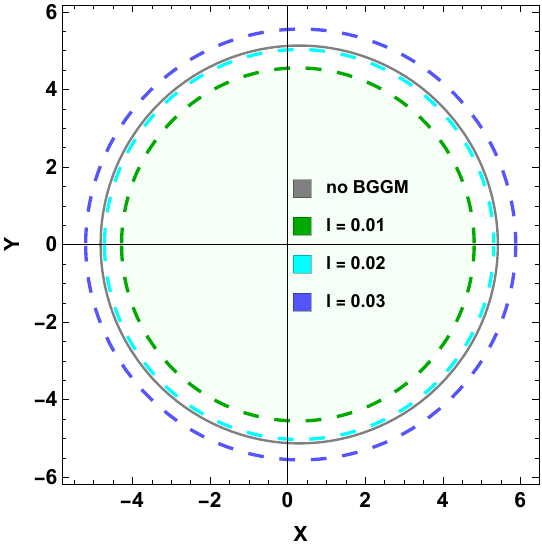}
	\hfill
	\includegraphics[width=.24\textwidth]{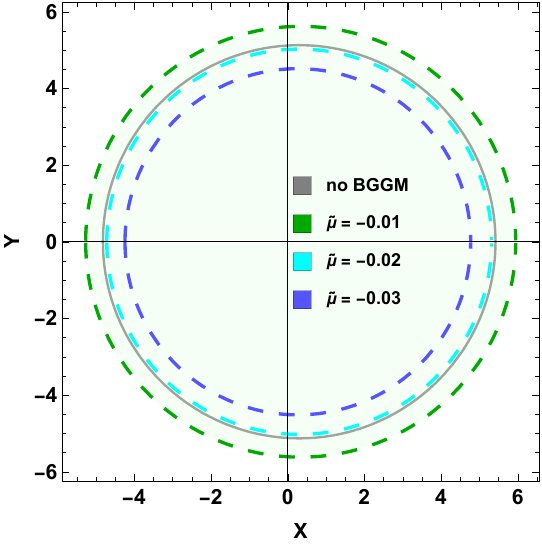}
	\hfill
	\includegraphics[width=.24\textwidth]{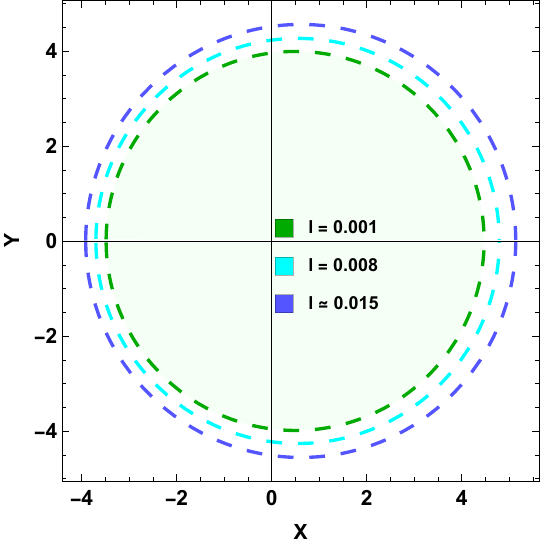}
	\hfill
	\includegraphics[width=.24\textwidth]{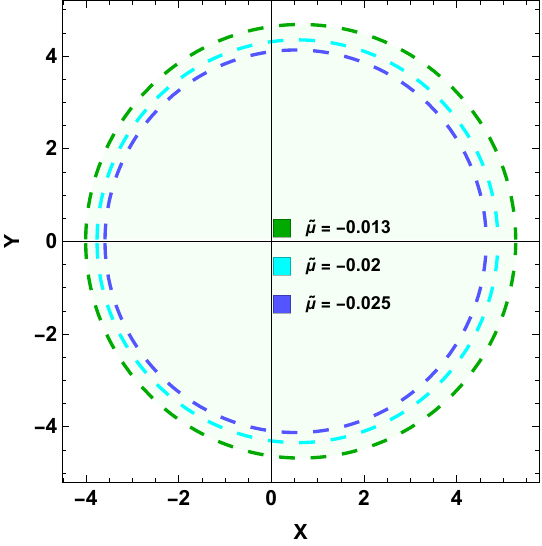}
	\hfill
	\includegraphics[width=.24\textwidth]{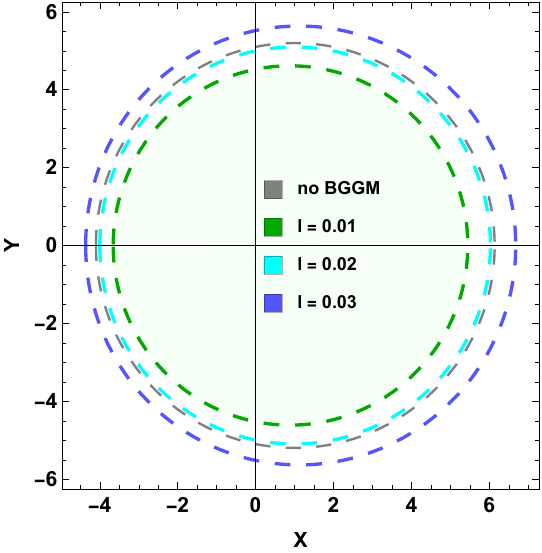}
	\hfill
	\includegraphics[width=.24\textwidth]{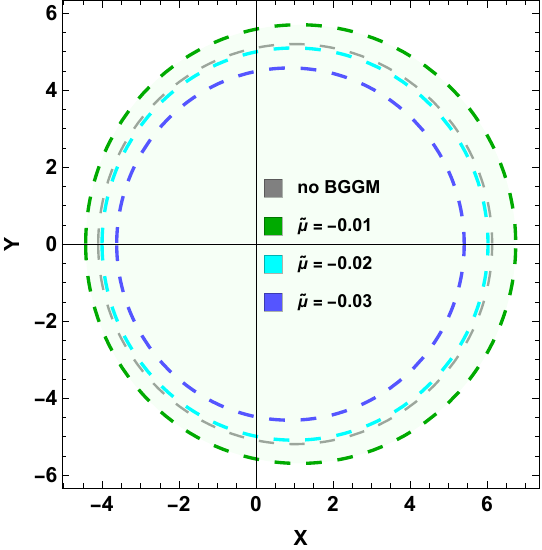}
	\hfill
	\includegraphics[width=.24\textwidth]{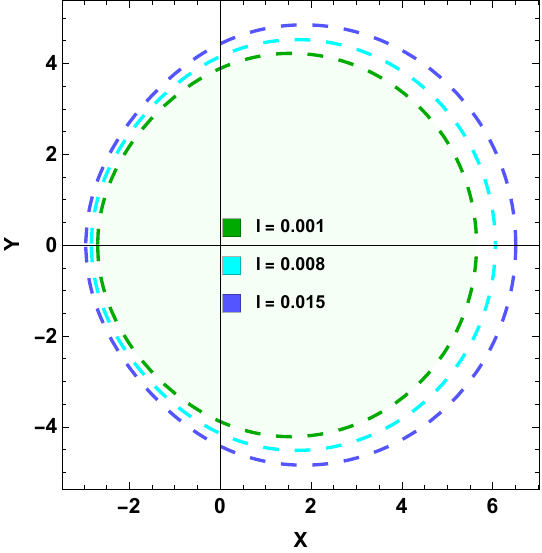}
	\hfill
	\includegraphics[width=.24\textwidth]{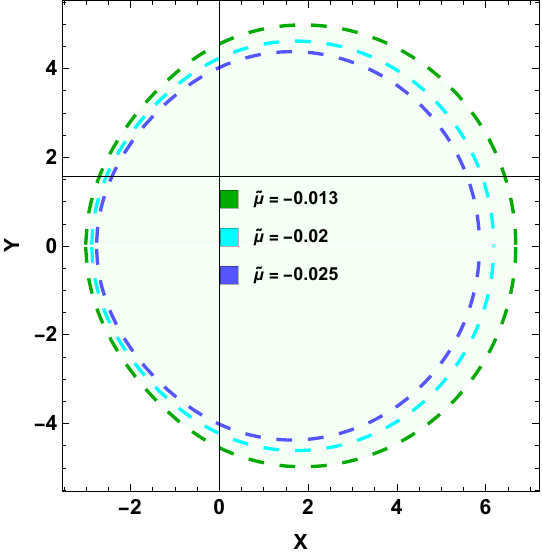}
	\caption{\label{fig:Sec52} 
		Shadow silhouette of the BGMKLBH, varying the parameter space $(a, \tilde{\mu}, \ell)$ at inclination angles $\theta_0 = 17^\circ$ (top row) and $\theta_0 = 90^\circ$ (bottom row). First column: Shadow shape variations with $a = 0.5$, $\tilde{\mu} = -0.02$, and variable $\ell$. Second column: Shadow shape variations with $a = 0.5$, $\ell = 0.02$, and variable $\tilde{\mu}$. Third column: Shadow shape changes with $a = 0.98$, $\tilde{\mu} = -0.02$, and variable $\ell$. Fourth column: Shadow shape changes with $a = 0.98$, $\ell = 0.01$, and variable $\tilde{\mu}$. Here, we set $M_{\rm{BH}} = 1$, $r_{\rm{b}} = 2R_{\rm{s}}$, and the observer located at $r_{\rm{O}} = 2.8 \times 10^{10} R_{\rm{s}}$.
	}
\end{figure*}
The shadows of the BGMKLBH scenario for the $\text{M87}^{*}$ SMBH, shown in Figure~\ref{fig:Sec52}, reveal intriguing details by taking realistic parameters for the DM spike, allowing us to observe how these shadows change with variations in the parameter space $(a, \tilde{\mu}, \ell)$. Regardless of the presence of the BGGM effect, an increase in DM density ($\rho_{\text{sp}}$) is expected to gradually expand the shadow radius \cite{NampalliwarAJ2021}. The shadow size is significantly influenced when the BGGM parameters $(\tilde{\mu}, \ell)$ vary by an order of $10^{-3}$.
The upper panels of \ref{fig:Sec52}, with $\theta_{0} = 17^{\circ}$, demonstrate that the shadow's boundary remains nearly circular even at extreme spin parameters, significantly influenced by variations in $\tilde{\mu}$ and $\ell$ on the order of $10^{-3}$. In contrast, the geometry of the BGMKLBH shadow exhibits significant changes for inclination angles exceeding \(86^{\circ}\)-- a common theoretical choice -- as illustrated in the lower panels of Figure~\ref{fig:Sec52}, where $\theta_{0} = 90^{\circ}$.
It is observed that when the influence of the BGGM diminishes, the BH shadow size approaches the Schwarzschild shadow radius of $3\sqrt{3}M$. Indeed, the shadow size decreases monotonically with reduced values of $\tilde{\mu}$ and $\ell$. As shown in the third and fourth columns of Figure~\ref{fig:Sec52}, the shadow deformation becomes more pronounced as the spin parameter $a$ approaches $1$. Unlike the standard Kerr case, where the shadow deformation manifests as an indentation on the left side due to the Lense-Thirring effect \cite{JohannsenApJ2010}, here we observe a distinct deformation, that is, the left side of the shadow bulges outward, moving away from the center, rather than exhibiting the typical indentation characteristic of the Kerr scenario. 
We also observe a horizontal shift in the shadow's center along the positive direction of the $X$-axis, which depends on increasing the parameter space $(a, \ell, \tilde{\mu})$. 
One notable characteristic of the shadow of a BH surrounded by a DM spike and influenced by the BGGM effect is its tendency for the left edge of the shadow to shift leftward, while the shadow?s center shifts to the right. This shift becomes more pronounced as the parameter space $(a, \tilde{\mu}, \ell)$ increases, by varying one parameter while holding the others constant. This phenomenon is not observed in the case of Kerr BHs.  
Motivated by these distinctive shadow features of BGMKLBH, we will use the shadow radius to impose constraints on the BGGM parameters.
To achieve the goal of plotting Figure \ref{fig:Sec52}, we followed the definition provided in Ref. \cite{FengEPJC2020}, where the typical shadow radius is established by considering the leftmost and rightmost coordinates relative to $r_{\text{ph}}^{-}$ and $r_{\text{ph}}^{+}$. This radius is given by
\begin{equation}\label{ShadowRadius}
	R_{\text{sh}}=\frac{1}{2}\left(X\left(r_{\text{ph}}^{+}\right)-X\left(r_{\text{ph}}^{-}\right)\right). 
\end{equation}
For general values of $\theta_{0}$ (except when viewing the shadow from the north pole, $\theta_{0}=0$, or the equivalent south pole, $\theta_{0}=\pi$), we can obtain $r_{\text{ph}}^{\pm}$ by solving the equation $Y(r=r_{\text{ph}}^{\pm},\theta_{0})=0$.

\section{Constraints from the EHT Observations
	\label{sec6}}
The shadow of BHs, which embodies the characteristics of the background spacetime in its distinctive shape and size, can be employed as a valuable tool for examining the fundamental theories of gravity and cosmology. Additionally, it can be employed to constrain the deviation parameters within these theories.
In this section, our motivation is to assess the validity of the BGGM model in the context of a deformed Kerr BH surrounded by the DM spike, using of the EHT results within the strong-field regime \cite{EventHorizonL1,EventHorizonL2,EventHorizonL3,EventHorizonL4,EventHorizonL5,EventHorizonL6,EventHorizonL12,EventHorizonL13,EventHorizonL14,EventHorizonL15,EventHorizonL16,EventHorizonL17,DoSci2019,GRAVITYcollaboration,VagnozziCQG2023, UniyalPoDU2023,KhodadiPRD2022,XavierPRD2023}. To achieve this goal, it is necessary to do further investigation into the BGMKLBH spacetime. Therefore, applying shadow observable, such as shadow radius/angular diameter, we are able to make an estimation of the BGGM parameters $\tilde{\mu}$ and $\ell$.
In this approach, we consider $\text{M87}^{*}$ \cite{EventHorizonL1,EventHorizonL2,EventHorizonL3,EventHorizonL4,EventHorizonL5,EventHorizonL6} and $\text{Sgr A}^{*}$ \cite{EventHorizonL12,EventHorizonL13,EventHorizonL14,EventHorizonL15,EventHorizonL16,EventHorizonL17} SMBHs as BGMKLBHs. We put constraints on the BGGM parameters using EHT shadow observations for $\text{M87}^{*}$ and $\text{Sgr A}^{*}$, as summarized in Tables~\ref{TabConst1} and \ref{TabConst2}, at inclination angles of $17^{\circ}$ and $46^{\circ}$, respectively 
\cite{BanerjeeJCAP2022,AllahyariJCAP2020,KhodadiJCAP2020,MengPLB2023,MengPRD2022}. To achieve this, we exploit observational data associated with the SMBHs and the DM spike profile within the $\text{M87}$ and Milky Way galaxies.
To characterize the BGMKLBH shadows and estimate the BGGM parameters, we examine two shadow observables: the shadow radius and the angular diameter. The shadow radius allows us to impose constraints on the BGGM parameters. 

\begin{table}[htb]
	\caption{
		Acceptable values for $\ell$ and $\tilde{\mu}$ can be determined based on three different values of $a = 0.5, 0.75,$ and $0.94$, depicted in Fig. \ref{fig:Sec56} by cyan, green, and blue lines, respectively. These values correspond to the BH shadow radius that matches the EHT horizon-scale image of $\text{M87}^{*}$  within the $1\sigma$ and $2\sigma$ confidence levels. \\ }
	\label{TabConst1}\centering
	\begin{tabular}{c|ccccc}
		\hline
		{$\ell$} &\multicolumn{2}{c}{$1\sigma$}& & \multicolumn{2}{c}{$2\sigma$} \\
		\cline{2-3} \cline{5-6}
		
		{} & {Lower} & {Upper} &{}& {Lower} & {Upper} \\
		\hline
		$a = 0.5$\,\,\, & $0.0034$ & $0.0384$ & & $--$ & $0.0519$ \\
		$a = 0.75$ & $0.0054$ & $--$ & & $--$ & $--$ \\
		$a = 0.94$ & $0.0078$ & $--$ & & $--$ & $--$ \\
		\hline
		{$\tilde{\mu}$} &\multicolumn{2}{c}{$1\sigma$}& & \multicolumn{2}{c}{$2\sigma$} \\
		\cline{2-3} \cline{5-6}
		
		{} & {Upper} & {Lower} &{}& {Upper} & {Lower} \\
		\hline
		$a = 0.5$\,\,\, & $-0.0037$ & $-0.0355$ & & $--$ & $-0.0602$ \\
		$a = 0.75$ & $--$ & $-0.0335$ & & $--$ & $-0.0582$ \\
		$a = 0.94$ & $--$ & $-0.0312$ & & $--$ & $-0.0560$ \\
		\hline
	\end{tabular}
\end{table}

\begin{table}[htb]
	\caption{
		Acceptable values for $\ell$ and $\tilde{\mu}$ can be determined based on three different values of $a = 0.5, 0.75,$ and $0.94$, depicted in Fig. \ref{fig:Sec56} by cyan, green, and blue lines, respectively. These values correspond to the BH shadow radius that matches the EHT horizon-scale image of $\text{Sgr A}^{*}$ within the $1\sigma$ and $2\sigma$ confidence levels. \\ }
	\label{TabConst2}\centering
	\begin{tabular}{c|ccccc}
		\hline
		{$\ell$} &\multicolumn{2}{c}{$1\sigma$}& & \multicolumn{2}{c}{$2\sigma$} \\
		\cline{2-3} \cline{5-6}
		
		{} & {Lower} & {Upper} &{}& {Lower} & {Upper} \\
		\hline
		$a = 0.5$\,\,\, & $0.0096$ & $0.0239$ & & $0.0015$ & $0.0304$ \\
		$a = 0.75$ & $0.0115$ & $--$ & & $0.0030$ & $--$ \\
		$a = 0.94$ & $0.0125$ & $--$ & & $0.0039$ & $--$ \\
		\hline
		{$\tilde{\mu}$} &\multicolumn{2}{c}{$1\sigma$}& & \multicolumn{2}{c}{$2\sigma$} \\
		\cline{2-3} \cline{5-6}
		
		{} & {Upper} & {Lower} &{}& {Upper} & {Lower} \\
		\hline
		$a = 0.5$\,\,\, & $-0.0165$ & $-0.0295$ & & $-0.0107$ & $-0.0373$ \\
		$a = 0.75$ & $--$ & $-0.0278$ & & $--$ & $-0.0357$ \\
		$a = 0.94$ & $--$ & $-0.0268$ & & $--$ & $-0.0349$ \\
		\hline
	\end{tabular}
\end{table}

As shown in Refs. \cite{NampalliwarAJ2021,XuJCAP2018-2,CapozzielloJCAP2023,HouJCAP2018,JusufiEPJC2020354,JusufiPRD2019,JusufiMNRA2021,PantigJCAP2022}, the shadow radius increases gradually with increasing DM density; nonetheless, for realistic DM settings, the BH shadows in our scenario remain nearly unchanged from what they would be in the absence of DM distribution. 
Nonetheless, the shadow silhouette is quite sensitive to the value of the model parameters, particularly in the case where our Kerr-DMS BHs behave as an extreme or near-extreme Kerr BH, with a spin parameter of $a\rightarrow 1$.

Refs. \cite{EventHorizonL1,EventHorizonL6} reported that the mass of M87* is $M_{\text{M87*}} = (6.5 \pm 0.7) \times 10^{9} \,\text{M}_{\odot}$, with an angular diameter $\theta_{\text{M87*}}= 42 \pm 3 \,\mu\text{as}$ of the BH shadow, as well as a distance from Earth of $D_{\text{M87*}} = 16.8\pm0.8\,\text{Mpc}$. 
Taking into account the Schwarzschild shadow deviations $\delta_{\text{M87*}} = -0.01\pm 0.17 $, the relation $\frac{R_{\text{Sh}}}{M} = 3\sqrt{3}(1+\delta_{\text{M87*}})$ gives that the shadow radius of  M87*  is restricted to the range $[4.26,6.03]$ in the \( 1\sigma \) confidence levels (CLs). 
As mentioned in Ref. \cite{EventHorizonL12}, results of Sgr A* show an angular diameter $\theta_{\text{Sgr A*}} = 48.7 \pm 7 \, \mu\text{as}$ of the BH shadow. The inferred distance from Sgr A* to Earth is given as $D_{\text{Sgr A*}} = 8277\pm 9 \pm 33\, \text{pc}$ (VLTI, standing for ``Very Large Telescope Interferometer''), $7953\pm50\pm32\, \text{pc}\, \text{(Keck)} $, with the BH mass $ M_{\text{Sgr A*}} = (4.297\pm 0.012 \pm 0.040) \times 10^{6} \, \text{M}_\odot \, \text{(VLTI)},\, (3.951\pm 0.047) \times 10^{6} \, \text{M}_\odot \, \text{(Keck)},\, (4.0^{+1.1}_{-0.6}) \times 10^{6} \, \text{M}_\odot \, \text{(EHT)}$. Based on Keck and VLTI measurements, the fractional deviation from the expected Schwarzschild values for Sgr A* is reported as $ \delta_{\text{Sgr A*}} = -0.08^{+0.09}_{-0.09} \, \text{(VLTI)} $ and $ \delta_{\text{Sgr A*}} = -0.04^{+0.09}_{-0.10} \, \text{(Keck)} $ \cite{EventHorizonL17,DoSci2019,GRAVITYcollaboration}. Taking the average of the Keck and VLTI estimates, represented as $\delta_{\text{Sgr A*}} \simeq 0.060^{+0.065}_{-0.065} \, \text{(Avg)}$, and applying $\frac{R_{\text{Sh}}}{M} = 3\sqrt{3}(1 + \delta_{\text{Sgr A*}})$ to define the shadow radius level, the size of the Sgr A* shadow is constrained within the range \([4.55, 5.22]\) in the $1\sigma$ CLs.
These derived bounds aim to constrain the deviations of the BGMKLBHs from Kerr BHs.
For the current analysis, we use the observational data for the SMBHs and their associated galaxies to the DM spike profile, as detailed in Section 3.
Using the observable $R_{\text{sh}}$, the shadow observable, namely the angular diameter of the  $\text{M}87^{*}$ and $\text{Sgr A}^{*}$ SMBHs, can be determined as 
\begin{equation}\label{ShadowAngDiam1}
	2R_\mathrm{sh} = \frac{\theta_{\text{sh}}D_{O}}{M_{\mathrm{BH}}}.
\end{equation}
According to Eq. \eqref{ShadowAngDiam1}, the shadow radii of the 
$\text{M}87^{*}$ and $\text{Sgr.A}^{*}$ that have been identified is strikingly compatible with the Schwarzschild BH surrounded by the realistic DM spike distribution.
This can be verified via setting the theoretical shadow diameter as $d^{\text{theo}}_{\text{sh}} = 2R_{\text{sh}}$. 
Then, Eq. \eqref{ShadowAngDiam1} can also be rewritten as
\begin{equation}\label{ShadowAngDiam2}
	\theta_{\text{sh}}=2\times9.87098\times 10^{-6}R_{\mathrm{sh}}\frac{M_{\mathrm{BH}}}{M_{\odot}}\frac{1\,\text{kpc}}{D_{O}}\,\mu \text{as} .
\end{equation}
{The first and third columns of Figure~\ref{fig:Sec56} display how the shadow radius varies with model parameters $\ell$ and $\tilde{\mu}$ for $\text{M87}^*$ and $\text{Sgr A}^*$. It illustrates how the EHT-derived allowed shadow radius region constrains $\ell$ and $\tilde{\mu}$ at $1\sigma$ and $2\sigma$ CLs as $a$ varies from $0.5$ to $94$, providing lower and upper bounds for these parameters while varying one and holding the others constant. 
	The numerical bounds for $\ell$ and $\tilde{\mu}$ are provided in Tables~\ref{TabConst1} and \ref{TabConst2}. No upper bounds were observed for the BGGM parameters when $a$ was fixed at $0.75$ and $0.94$.	
	Thus, it not only offers the critical bounds for $\ell$ and $\tilde{\mu}$, but also indicates how the shadow radius evolves as $\ell$ (or $\tilde{\mu}$) changes, while keeping the observer?s radial distance $r_{O} \equiv D_{O}$ from the black hole fixed.	
	
	\begin{figure*}[htb]
		\centering 
		\includegraphics[width=.24\textwidth]{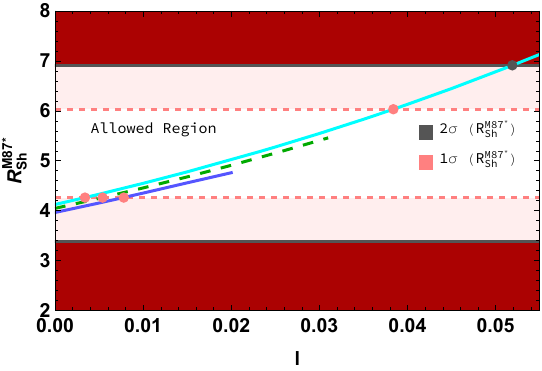}
		\hfill
		\includegraphics[width=.24\textwidth]{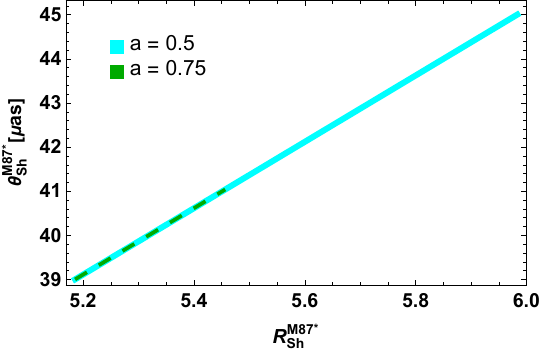}
		\hfill
		\includegraphics[width=.24\textwidth]{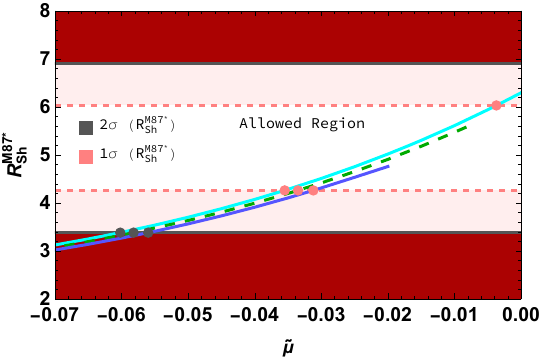}
		\hfill
		\includegraphics[width=.24\textwidth]{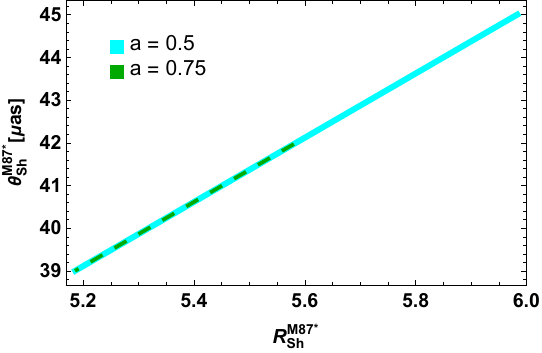}
		\hfill
		\includegraphics[width=.24\textwidth]{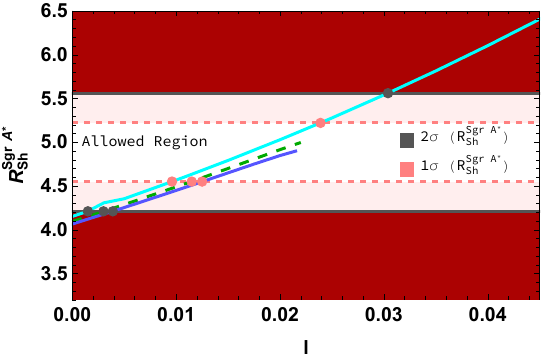}
		\hfill
		\includegraphics[width=.24\textwidth]{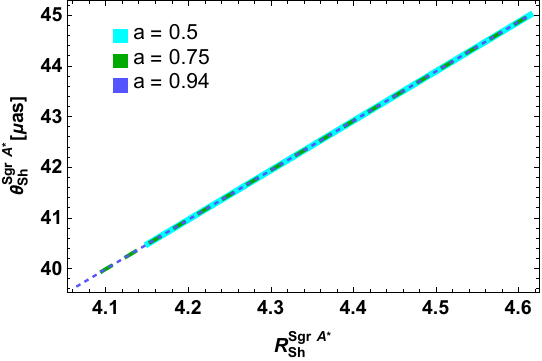}
		\hfill
		\includegraphics[width=.24\textwidth]{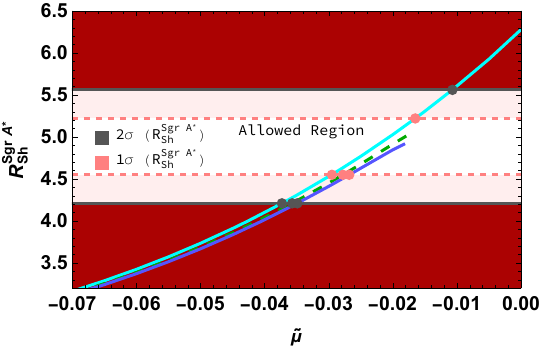}
		\hfill
		\includegraphics[width=.24\textwidth]{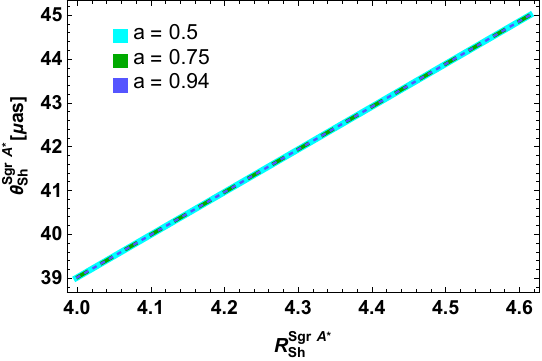}
		\caption{\label{fig:Sec56} 
			Figure depicting the shadow radius and angular diameter dependencies on $\ell$, $\tilde{\mu}$ and the observable $R_{\rm{sh}}$, respectively, for three distinct $a$ values at an inclination angle $\theta_{0}=17^{\circ}$.
			First column: BGMKLBH shadow radius plotted against parameter $\ell$ for $\tilde{\mu} = -0.02$ with three values of $a$ ($0.5,\,0.75,\,0.94$), represented by cyan, green (dashed), and blue curves, respectively. Third column: shadow radius vs. $\tilde{\mu}$ for $\ell = 0.02$ with the same $a$ values and color scheme. The BH mass is set to $M_{\rm{BH}} = 1$.  Second and fourth columns:
			angular diameter behavior for BGMKLBH in terms of the observable $R_{\rm{sh}}$, showing dependencies on $\ell$ (with $\tilde{\mu} = -0.02$ and $a = 0.5, 0.75$, in the second column), and on $\tilde{\mu}$ (with $\ell = 0.02$ and $a = 0.5, 0.75, 0.94$, in the fourth column, with dotted blue for $a=0.94$). Red shaded regions highlight values of $\ell$ and $\tilde{\mu}$ inconsistent with stellar dynamics observations for $\text{M}87^{*}$ (top row) and $\text{Sgr A}^{*}$ (bottom row). White and light pink areas represent ranges consistent with EHT horizon-scale images for $\text{M}87^{*}$ and $\text{Sgr A}^{*}$ at $1\sigma$ and $2\sigma$ CLs, while in the bottom row, these shaded regions align with Keck and VLTI mass-to-distance ratio priors for $\text{Sgr A}^{*}$.
		}
	\end{figure*}
	Besides, Figure~\ref{fig:Sec56} presents the connection between the shadow radius and angular diameter, respectively, with respect to the parameter space $(a,\ell,\tilde{\mu})$ and the observable $R_{\text{sh}}$ for SMBHs $\text{M}87^{*}$ and $\text{Sgr A}^{*}$. 
	In the second and fourth columns of Figure~\ref{fig:Sec56}, we depict the angular diameter \eqref{ShadowAngDiam2} versus $R_{\text{sh}}$
	for $\text{M}87^{*}$ and $\text{Sgr A}^{*}$, respectively, as a function of $\tilde{\mu}$ and $\ell$ for three distinct $a$ values.
	In this figure, the angular diameter is calculated using previously determined values for the distance, with $D_{O} = 16.8\, \text{Mpc}$ and mass $M_{\text{M}87^{*}} = 6.4 \times 10^9 M_{\odot}$ for $\text{M}87^{*}$, and $D_{O} = 8.3\, \text{kpc}$ and mass $M_{\text{Sgr.A}^{*}} = 4.1 \times 10^6 M_{\odot}$ for $\text{Sgr A}^{*}$. 
	In standard assumptions, the DM profile is commonly considered to disappear below $4R_{s}$ \cite{GondoloPRL1999}, or $2R_{s}$ according to full relativistic calculations for a static BH \cite{SadeghianPRD2013}, due to the BH capturing DM particles. To simplify, when constraining the model parameters $\tilde{\mu}$ and $\ell$ using EHT observations of $\text{M}87^{*}$ and $\text{Sgr A}^{*}$ shadows, assuming a power-law density for the DM spike near these SMBHs, we adopt a DM spike extending all the way down to $2R_{s}$ for the BGMKLBHs. However, this simplification has a minimal impact on our results.
	
	In this way, we infer that the EHT observations of the $\text{M}87^{*}$ and $\text{Sgr A}^{*}$ BH shadows do not exclude the rotating Kerr-like BH within the BGGM gravity model.
	By modeling the rotating BGGM-BH surrounded by a DM spike as the SMBHs $\text{M87}^*$ and $\text{Sgr A}^*$, we report the most stringent constraints on the BGGM parameters at $a \leq 0.5$.
	Specifically, the curves in Figure~\ref{fig:Sec56} demonstrate that the spin parameter strongly influences the constraints on the model parameters. Thus, we fix $a = 0.5$ and plot the CLs, showing the upper limits for $\ell$ and $\tilde{\mu}$ from the EHT observations. At the $95\%$ CLs, the upper limits are $\ell \leq 0.0384, \tilde{\mu} \leq -0.0037$ for $\text{M}87^{*}$ and $\ell \leq 0.0239, \tilde{\mu} \leq -0.0165$ for $\text{Sgr A}^{*}$. At the $68\%$ CLs, the upper limits are $\ell \leq 0.0519, \tilde{\mu} \leq \text{None}$ for $\text{M}87^{*}$ and $\ell \leq 0.0304, \tilde{\mu} \leq -0.0107$ for $\text{Sgr A}^{*}$ \cite{KuangAP2022,AnjumPoDU2023}.
	From Tables~\ref{TabConst1} and \ref{TabConst2}, we infer that the astrophysical constraints on the parameters $\ell$ and $\tilde{\mu}$ indicate that $\text{Sgr A}^{*}$ yields stronger constraints than those derived for $\text{M}87^{*}$.
	Given the current precision of astrophysical observations, we find that $\text{M}87^{*}$ and $\text{Sgr A}^{*}$ could be BGMKLBHs.

	
	
	\section{Discussion and Conclusions
		\label{Conc}
	}

A unique opportunity to test the strong-field predictions of GR and shed light on metric theories of gravity is provided by the SMBHs, $\text{M}87^{*}$ and $\text{Sgr A}^{*}$, at the core of the M87 and Milky Way galaxies.
Although facing certain theoretical challenges, these two compact entities appear to be highly plausible candidates for representing astrophysical BHs.
Images of the compact objects at the Galactic Centers obtained by the EHT collaboration inspired us to investigate the properties of BHs in a novel configuration: the BG combined with a GM in the region impacted by the DM spike. 
As we have previously stated, topological defects such as GMs may emerge in Lorentz violating theories, where the bumblebee field might effectively cause Lorentz violation. If  isolated, GMs might persist in the Universe up today.

We assume that the central regions, including the SMBHs themselves, are immersed in a DM spike characterized by a power-law density profile. 
In the DM spike-affected region, we investigate the combined impact of BG and GM effects, particularly focusing on the influence of parameters $\ell$ and $\tilde{\mu}$ on Kerr BH features like the horizons, ergoregions, SLSs, and shadows. 
Thus, the combination of both these components -- the BGGM effect and the DM spike profile -- may present a realistic platform for exploring the characteristics of the Kerr BH. It should be mentioned that no prior study has been done on a configuration containing these concepts. Therefore, we expect that this study will make a substantial contribution to our understanding of the effects of the bumblebee field, GM, and DM spike on the horizons, SLSs, ergoregions, and shadow silhouettes. 
For this aim, we estimated the normalization parameter, $\rho_{0}$, and associated parameters for the DM spike profiles in both the Milky Way and $\text{M87}$ galaxy centers.

The scarcity of rotating BHe models within a DM spike, influenced by BGGM effects, hampers progress in testing the BGGM model using observations, such as the EHT results of $\text{M}87^{*}$ and $\text{Sgr A}^{*}$.
To tackle this challenge, we commenced our study by considering a spherically symmetric, static BH with BGGM properties as the seed metric. From there, we develop a non-rotating spacetime incorporating a DM spike resulting in the emergence of the BGMSLBH spacetime.
To create the BGMSLBH spacetime, we began with a power-law density profile and solved the modified TOV equation, approximating the integral in the leading order for the spike density. Our approach involved the critical condition of matching the inner BH spacetime with the outer region, specifically employing the condition denoted as ${\rm{f}}(r_{\text{b}}) = e^{2\chi(r_{\text{b}})} = 1-\frac{2M_{\text{BH}}}{r_{\text{b}}}$, in line with the methodology presented by Nampalliwar et al. (2021). This process resulted in the calculation of corresponding metric components, where $\text{f}(r)\neq\text{g}(r)$.

We applied the modified NJ algorithm to generalize our approach to the rotating scenario, resulting in BGMKLBH spacetime metrics with a Kerr-like form. EHT results are in line with Kerr metric predictions, with no evidence of GR violations. We then explored BGMKLBH, examining horizons, SLSs, ergoregions, and shadow images. Once the BGGM effects are turned off (i.e., $\ell = 0 = \tilde{\mu}$) and the DM spike vanishes (i.e., $\rho_{\text{sp}} \rightarrow 0$ or $r \rightarrow r_{\text{b}}$), the BGMKLBH solutions reduce to the standard Kerr solution.

By modeling BGMKLBHs as $\text{M}87^{*}$ and $\text{Sgr A}^{*}$ SMBHs and using the estimated BH mass and distances, we demonstrated that various BH characteristics, such as the event horizon, outer SLS, shadow radius, and angular diameter remain almost unchanged compared to their counterparts without DM distribution for realistic DM parameters. 
However, we expect that a rise in the DM spike density would have a significant impact on these BH characteristics' sizes. Our findings indicated that, based on the existing observational data regarding DM spike density, the influence of DM is minimal. In our case, the modification introduced by BGGM is distinctive and discernible from the effects of BH spin and the surrounding DM spike.

Through numerical analysis of the event horizon ($r_+$), outer SLS ($r_{\text{SLS}}^+$), and ergoregion ($\delta_{\text{er}}$), we found that, aside from the parameters related to BH and the DM spike masses, changes occur in $r_{+}$, $r_{\text{SLS}}^{+}$ and $\delta_{\text{er}}$ as $a$, $\ell$ and $\tilde{\mu}$ vary. Specifically, as $\ell$ increases and $\tilde{\mu}$ decreases individually, with $a$ held constant, $r_{+}$ and $r_{\text{SLS}}^{+}$ reduce, while $r_{-}$ and $\delta_{\text{er}}$ expand. Likewise, with fixed $\ell$ and $\tilde{\mu}$, raising $a$ decreases $r_{+}$ and $r_{\text{SLS}}^{+}$ and increases $r_{-}$ and $\delta_{\text{er}}$.
To be more precise, for fixed values of the DM spike parameter and spin $a$, the GM effect has a greater influence in this scenario than the BG effect.

In our shadow analysis, we addressed photon geodesic equations, which we solved analytically in a first-order differential form.
In more realistic DM spike scenarios, we observe that even at extreme spin values, the BGMKLBH shadows with an inclination angle of $\theta_{0}=17^{\circ}$ maintain a nearly circular boundary, with minimal impact from the parameters $\ell$ and $\tilde{\mu}$ on the distortion of the BH shadow.
However, a significant change in the geometry of the BGMKLBH shadow is evident at an inclination angle of $\theta_{0} = 90^{\circ}$.
It is observed that as the BG and GM influences weaken, i.e., when $\ell$ and $\tilde{\mu}$ tend to zero, the BH shadow shrinks and expands, respectively.

As such, the shadow shape is quite sensitive to the BGGM parameters value, particularly in the case where the BGMKLBH exhibits characteristics of an extreme or near-extreme modified Kerr BH, with a spin parameter of $a\rightarrow 1$, in such a case, at the inclination angle of $\theta_{0} = 90^{\circ}$, a significant impact of the $\ell$ and $\tilde{\mu}$ parameters on the distortion of the BH shadow is observed.

{
	In the case of the BGMKLBH, the shadow deformation differs from the standard Kerr scenario. Instead of the typical indentation on the left side caused by the Lense-Thirring effect, the left edge of the shadow exhibits a bulge outward, away from the center. Additionally, a horizontal shift of the shadow?s center along the positive $X$-axis is observed, which depends on increasing the parameter space $(a, \ell, \tilde{\mu})$. A notable feature is that as the parameter values increase, the left edge of the shadow shifts leftward while the center moves rightward, behavior is not observed in Kerr BHs. 
	Motivated by the distinct shadow features of BGMKLBH, we employ the shadow radius to constrain the BGGM parameters.
	The shadow observables, including the shadow radius and angular diameter, were employed to quantify shadow size, enabling the estimation of BGGM parameters and the exploration of the BGMKLBH solution.
	As such, we considered $\text{M87}^{*}$ and $\text{Sgr A}^{*}$ SMBHs as BGMKLBHs. We put constraints on the BGGM parameters, using EHT shadow observations for $\text{M87}^{*}$ and $\text{Sgr A}^{*}$ at inclination angles of $17^{\circ}$ and $46^{\circ}$, respectively. 
	To obtain admissible values for the model parameters, we utilized available observational data associated with the SMBHs and the DM spike profiles within the $\text{M87}$ and Milky Way galaxies and derived constraints on the shadow radius from the EHT data.
	Observations of $\text{Sgr A}^{*}$ constrained the parameters within a more confined range, whereas $\text{M87}^{*}$ suggested a wider one. Notably, the $\text{Sgr A}^{*}$ data impose more robust constraints on the model parameters, including values extending beyond the upper $2\sigma$ CLs, compared to those for $\text{M87}^{*}$.
	Within a consistent parameter space for a modified Kerr BH surrounded by a DM spike and subject to the BGGM effect, EHT observations do not rule out the BGGM influence at galactic centers. Therefore, BGMKLBHs remain viable candidates for astrophysical BHs.
}
\\

\textbf{CRediT authorship contribution statement}\\
All the authors have the same contribution in preparing the paper. \\

\textbf{Declaration of competing interest}\\
There is no conflict of interest.\\

\textbf{Data availability}\\
No data was used for the research described in the article.

\section*{Acknowledgements}
The research of LMN, SZ and HH was supported by the Q-CAYLE project, funded by the European Union-Next Generation UE/MICIU/Plan de Recuperacion, Transformacion y Resiliencia/Junta de Castilla y Leon (PRTRC17.11), and also by project PID2023-148409NB-I00, funded by MICIU/AEI/ 10.13039/501100011033. Financial support of The Department of Education of the Junta de Castilla y Leon and FEDER Funds is also gratefully acknowledged (Reference: CLU-2023-1-05). 
This paper is partly based upon work from COST Action CA21136 {\it Addressing observational tensions in cosmology with systematics and fundamental physics} (CosmoVerse) supported by COST (European Cooperation in Science and Technology). SC  acknowledges the  {\it Istituto Nazionale di Fisica Nucleare}  ({\it iniziative specifiche} QGSKY and MOONLIGHT2).

\end{document}